\documentclass[12pt]{article}

\usepackage[many]{tcolorbox}

\definecolor{main}{HTML}{5989cf}    
\definecolor{sub}{HTML}{cde4ff}     

\usepackage{amsfonts,booktabs,epsfig}
\usepackage{mathtools}
\usepackage{lineno,hyperref}
\usepackage{graphicx}
\usepackage{enumerate}
\usepackage{natbib}
\usepackage{url}
\usepackage[utf8]{inputenc}
\usepackage{multirow}
\usepackage{latexsym}
\usepackage{enumitem}
\usepackage{xcolor}
\usepackage{dsfont}
\usepackage{mathrsfs}
\usepackage{verbatim}
\usepackage{float}
\usepackage{soul}
\newcommand{\PP}{\mathbb{P}\/}

\newcommand{\T}{\mathbb{T}}

\def\id{\hbox{{\rm\textbf 1}\kern-.5em\hbox{{\rm\textbf 1}}}\! }

\newcommand{\bi}{\begin{itemize}}
	\newcommand{\ei}{\end{itemize}}
\newcommand{\beq}{\begin{equation}}
	\newcommand{\eeq}{\end{equation}}

\newtheorem{dfn}{Definition}
\newtheorem{rmk}{Remark}

\newcounter {num} 

\newcounter {no} 

\newcounter {nrq} 

\begin{document}
	
		
		
		
		\title{A Hierarchical Decision-Based Maintenance for a Complex Modular System Driven by the {\it MoMA} Algorithm}

		\author{
		Mar{\'\i}a Luz G\'amiz \\ Department of Statistics and O.R., University of Granada, Spain\\ \\
		D. Montoro-Cazorla\\  Department of Statistics and O.R., University of Jaen, Spain\\ \\
		M.C. Segovia-Garc\'ia \thanks{Corresponding author: msegovia@ugr.es} 
		\\ Department of Statistics and O.R., University of Granada, Spain
	}	
	\maketitle
	     
		\begin{abstract}

		{This paper presents a maintenance policy for a modular system formed by $K$ independent modules (n-subsystems) subjected to environmental conditions (shocks). For the modeling of this complex system, the use of the Matrix-Analytical Method (MAM) is proposed under a layered approach according to its hierarchical structure. Thus, the operational state of the system (top layer) depends on the states of the modules (middle layer), which in turn depend on the states of their components (bottom layer). This allows a detailed description of the system operation to plan maintenance actions appropriately and optimally. We propose a hierarchical decision-based maintenance strategy with periodic inspections as follows: at the time of the inspection, the condition of the system is first evaluated. If intervention is necessary, the modules are then checked to make individual decisions based on their states, and so on. Replacement or repair will be carried out as appropriate. An optimization problem is formulated as a function of the length of the inspection period and the intervention cost incurred over the useful life of the system.  Our method shows the advantages, providing compact and implementable expressions. The model is illustrated on a submarine Electrical Control Unit (ECU).}

	\end{abstract}
	

	\noindent%
		
		

	
	
	\section{Introduction}
	\label{sec:int}
\subsection{Motivation}
Modularity is a concept widely applied in many fields, such as manufacturing industry, power and energy systems, computer science, smart grid, biology. In fact, many modern systems consist of modules e.g. aircrafts, rockets, submarines, automobiles, high-tech medical equipment and others. The advantages of using modular product design are discussed widely in the existing literature, which also shows a variety of recent applications to engineering systems \cite{Drossel2019}-\cite{Nilda2020}. When defining the modules, diverse strategies are found based on different motivations or design goals \cite{Jeremy2016},\cite{Baldwin2000}. Thus, a modular system can be defined according to either its functioning requirements \cite{Li2021}, manufacturing considerations \cite{Shaik2015}, reliability optimization \cite{Dorronsoro2022} or maintainability targets \cite{Yicong2016} among others. 

In practice, most engineering systems have to fulfill missions that require high reliability, so they need to be maintained regularly to avoid catastrophic failures and costly disruptions. In this respect, continuous system monitoring or inspections over time are key to decide whether a maintenance action should be performed, which could be preventive or corrective. Preventive maintenance (PM) actions attempt to maintain the system in an acceptable condition of operation by avoiding a system failure. Corrective maintenance (CM) refers to all actions that occur when the system has already failed. For some systems, continuous monitoring is not feasible and inspections turn out to be the only alternative tool. Thus, determining the optimal frequency of inspections to minimize its cost while keeping the system performance is essential. An extensive amount of literature deals with multi-component models under different approaches that involve periodic inspections, age-based maintenance or condition-based maintenance; where reliability and cost optimization problems are often formulated \cite{Sharifi2020}-\cite{Hashemi2020}. 

The present work focuses on multi-component systems designed with a modular structure. Units (components) within each module are assumed to be similar in some aspects, e.g. units  that perform a certain mission or function in the system \cite{Zhao2020},\cite{Li2021}, units that compose a redundant structure \cite{Li2016}, units making up voting circuits in electrical or computer systems \cite{Liu2014} and so on. In all cases, a module can be represented by a subsystem of components. Here, a system designed with $K$-independent modules (KMS) is studied, where each module is an $n_i$-subsystem, and the system performs its function through its modules. These  modules are assumed to operate under environmental conditions causing shocks that could affect each module in a different way. The modeling of the shocks is done by Markovian Arrival Processes {(MAPs)}. The lifetimes of the units within the modules are assumed to be {Phase-type distributed} (PH-distributed). {This system was already considered by the authors in} \cite{MOMA2022}, {where primarily a reliability analysis was performed, and, maintenance was only treated briefly. Specifically, the optimal length of the operating time of the system until it is inspected for the first time was determined based on the probability ($q$) that the system reaches a critical operating state. In} \cite{MOMA2022}{, certain cost-based arguments were used to plan the optimal inspection schedule, but no detailed analysis based on maintenance actions and cost specifications as it is done here was carried out.}

{It is clear that} modeling complex systems is a challenging task due to the large number of units, structures and factors concerned, which lead to an arduous and large dimensional framework. For this reason, many multi-component systems are studied in the literature using numerical or simulation techniques \cite{Chen2022}. {As an alternative}, Markovian processes and the use of the Matrix-Analytical Method(MAM) has proven to be a useful tool for the study of complex systems \cite{Montoro2015}-\cite{Chakravarthy2012}. {Nevertheless, to the authors' knowledge, reliability literature on modular systems is very limited and even more so using the above methodology. Thus, for modeling this type of system, we propose the use of the MAM under a layered/leveled approach according to its hierarchical structure where the whole system is located at the top, the modules are in the middle and the components are set at the bottom. As a first goal, in} \cite{MOMA2022} {the authors designed the so-called {\it MoMA} (Modular Matrix-Analytic Method) Algorithm to build (bottom-up) the generator of this system and from it the assessment of the system was carried out. However, the present work focuses on the design of a maintenance strategy for the system. To this end, our approach is most appropriate by allowing a state-based hierarchical description of the system operational conditions, i.e. states at each level capture the operational health (states) of every element at the level/s below. This provides detailed information that makes possible to plan maintenance actions in a precise and optimal way. Regarding the layered description of systems, other works can be found in the literature under different targets and methodologies. For instance, the work in} \cite{Torrado21} {deals with a multi-level redundancy assignment problem in a modular system, where the effect of redundancies (at different levels) on the reliability of the system is evaluated. For maintenance purposes for a multi-component system, the authors in} \cite{Chen2022} {propose a preventive and opportunistic method with a two-phased inspection: the system is first inspected and, depending on its condition, in the second phase, the units are inspected or not.}

{For our system, the states (at each level) can be categorized into groups according to operational health under progressive deterioration or risk, i.e. optimal, critical and failure. Thus,} we propose the following inspection-maintenance policy: at the time of the inspection, the operational state of the system is assessed first (optimal, critical or failure), which determines whether or not a maintenance action is required. If so, then it becomes necessary to look at the operational state of each module to make individual decisions for each. Those modules which are in a critical or failure state are ascertained. For the rest, no intervention will take place. Thus, maintenance tasks will be performed accordingly. As can be noted, this maintenance strategy aligns with system design: system, module and component levels can be distinguished. It is well justified for those multi-component systems where it is not feasible or costly to inspect every single component. For example, in a battery pack system in electric vehicles, voltage checking for the bank of cells is cheaper than for individual cells, taking into account the assembling and disassembling costs \cite{Chen2022}. {In connection with this, existing literature establishes that in models with positive economic dependencies maintaining similar units together is cheaper than maintaining them separately since the set-up costs can be shared as well as a specialized technicians team when similar maintenance tasks are required} \cite{Maaroufi2013},\cite{Levi2014}. {In addition, replacing an entire module when it fails is sometimes a more economical option than replacing each of its components} \cite{SeongJong2009},\cite{Uday1987}.

In short, we plan for the system a hierarchical and decision-based maintenance with periodic inspections, where preventive and corrective actions are included. These actions are implemented in blocks, i.e. several units could be repaired at the same time or jointly replaced. 
The maintenance effect on the modules is mathematically represented by matrices, which describe the transitions rules from the occupied state just before the inspection to the state (to be occupied) just after the inspection. These matrices play an important role when determining the Markovian process that governs the evolution of the system over time. As will be shown later, a  recursive procedure allows to describe the Markovian process along each inspection period starting from the first one, where the system has not yet been inspected (addressed in \cite{MOMA2022}). A cost analysis on each inspection period is performed, based on maintenance costs, e.g. per inspection, per replacement of each unit, per down-time, and others.  Finally, an optimization problem is formulated as a function of the time between inspections and the incurred costs along the useful life of the system, where a maximum number of inspections is allowed. The expressions and calculus are illustrated on a real system, consisting of the electrical control equipment of a submarine, formed by four modules. A discussion of the main results and conclusions are reported.

\subsection{Contributions}
{The main contributions of this paper are listed below}. 

\begin{enumerate}
	
	\item {Modules are key to system reliability and maintainability: with $K$ independent modules, a failure in one module generally does not compromise the entire system operation, although its performance might degrade. Thus, system malfunctions or failures can be identified and located more easily than in non-modular logical structures.  As a result, the system maintenance process is simplified, potentially reducing time and costs. Our approach takes these aspects of modular design theory into account.}. 	
	
	\item {The method we propose proves to be suitable for modeling this kind of complex modular systems. Specifically, it allows:}
	
	\begin{itemize}
		\item {To give a state-based hierarchical description of the system operational health, which provides detailed information. Thus, the state of the system collects information about the state of each of its elements, as each level captures the state of the elements at the level below. This allows for accurate assessments of the operational conditions at the time of inspection, which enables maintenance needs to be precisely established. This is an important contribution, when compared with those systems that behave like black boxes at the time of inspection-maintenance. }
		
		\item {To plan a hierarchical condition-based policy accordingly.}
		
		\item {To perform a stepwise analytical study, by levels, where mathematical tools such as MAPs, PH-distributions and Kronecker operators are essential and lead to compact and algorithmic expressions that make the model implementable. Specifically, matrices that are relevant in this maintenance analysis are novel and easily constructed under this approach.}
		
	\end{itemize}	
	
	\item {For the KMS described, already considered in} \cite{MOMA2022}, {a maintenance policy based on regular inspections along its useful life is provided. A cost minimization problem is formulated in terms of the frequency of inspection.}

	\item {PH-distributions and MAPs extend the Exponential distribution and Poisson arrival process, respectively, among others, often assumed in engineering studies. For example, the submarine ECU studied in} \cite{Liu2014} {assumes all lifetimes and repair times exponentially distributed; such study also reveals a complex framework due to the large number of states given for the system, which could be improved using the MAM, as we propose in this paper. Additionally, each module in the system is assumed to fail independently of the others based on two sources of failure: units wear-out and external shocks. This situation fits many modular systems in practice, where each module can be exposed to certain environmental or working conditions depending on the assigned function to perform in the system or depending on its physical location, e.g. modules underwater as considered in the paper referenced before.}
	
	\item {A case example concerning the Electrical Control Unit (ECU) of a submarine illustrates numerically, for different cost scenarios, the optimal length of the inspection period and, from this, the total number of inspections to be performed over the useful life of the system. }  
	
\end{enumerate}

Further contributions can be made by applying our method to the study of other complex models, such as the following ones: 1) systems that are decomposed into independent subsystems (modules) for simplification aims (see \cite{Qingzhu2022}); 2) modular systems where module is derived in another way, depending on the purpose or context of application. For example, a module can be interpreted as a group of components to be maintained in the same way (see \cite{Nilda2020}). In summary, our methodology is a powerful and flexible tool for the analysis of complex systems. Therefore, it is suitable for addressing a wide variety of reliability and maintainability problems taking advantage of the modular design. For example, by adding or updating functions, or adding redundancy in the  system by means of modules, the modeling of the system (after these changes) could be easily adapted. \\

The remainder of this paper is organized as follows. Section \ref{sec:preliminar} provides the relevant definitions for application of the MAM. Section \ref{sec:system} establishes the assumptions of the system. Sections \ref{sec:maintpolicy} and \ref{sec:cost} are devoted to the proposed maintenance policy for the system and the associated cost model. In these sections, the key mathematical elements for modeling the system maintenance and to formulate a cost optimization problem are defined. Section \ref{sec:application} illustrates the model on a real system, a submarine ECU comprising four modules. A discussion of the main results and  some conclusions are presented in Section \ref{sec:conclusions}.\\

\section{Preliminaries}\label{sec:preliminar}
In this section we present the nomenclature used and define some basic concept that will be needed for the rest of the paper.
\subsection{Acronyms $\&$ Notation}

	\begin{table}[H]
	\centering
	\begin{tabular}{|lll|}
		\hline 
		& & \\
		KMS &  & $K$-Modular System \\ 
		PH-distribution  &  & Phase-Type distribution \\ 
		MAP &  & Markovian Arrival Process \\ 
		MAM &  & Matrix-Analytical Method \\ 
		{\it MoMA}  &  & Modular Matrix-Analytic Alagorithm \\ 
		ECU &  & Electrical Control Unit \\ 
		SEM  &  & Subsea Electrical Module \\ 
		BOP &  & Blowout Preventer \\ 
	\hline
	\end{tabular}\caption{Acronyms}\label{tab:acron}
\end{table}

In Table \ref{tab:notation} we detail the notation used in the paper.
\begin{table}[H]
	\centering
{\small{
	\begin{tabular}{|lll|}
		\hline  
		$K$ &  & Number of modules in the system \\ 
		$n_{i}$ &  & Number of units in module $i=1, \ldots, K$ \\ 
		$\mathbb Q_{sys}$ &  & Infinitesimal generator of the system \\
		$U$&       & Set of operational states of the system \\ 
		$U_{1}$&   & Set of optimal states of the system\\
		$U_{2}$	&  & Set of critical states of the system  \\	
		$D$ &      &Set of down states of the system\\	
		$s_{ji}$ & & State of unit $j$ of module $i$ \\
		$\widetilde{s}_i$ & & state of module $i$ \\
		$\widetilde{\bf s}$ & & state of the system \\
		$M^j_{i}$& &Matrix of maintenance-effect on unit $j$ in module $i$\\
		$\beta_{j,i}$& &Vector of state-probability for unit $j$ in module $i$ after maintenance action \\
		${\bf M}_{i}$& &Matrix of maintenance-effect on module $i$\\
		$\beta_{i}$& &Initial probability vector for module $i$ after replacement of the module\\
		$\tau_{{\rm a}}$& & ${\rm a}th$ inspection time\\
		$A$	& &Maximum number of inspections\\
		$X(t)$ 	& &State of the system at time $t$, $0<t < \tau$ (without maintenance)\\
		$({\rm a}\tau, ({\rm a}+1)\tau]$ & & ${\rm a}$-th operational cycle of the system, ${\rm a}=0,\ldots,A-1$\\
		
		$\widetilde{\alpha}({\rm a})$& & System state probability vector after ${\rm a}$-th inspection\\
		$\widetilde{X}({\rm a},t)$& & State of the system at time ${\rm a} \tau+t$, $0<t < \tau$, ${\rm a}=1,\ldots,A$ \\
		$\widetilde{\alpha_{i}}({\rm a})$& &Vector of state probabilities for module $i$ after 
		${\rm a}$-th inspection. \\
		$\widetilde{X}_{i}({\rm a},\tau)$& & State of module $i$ at the end of the ${\rm a}$-th operational cycle \\
		$\beta$& & Initial probability vector of the system after full replacement\\
		${\bf C}_{i}$ & & Matrix of maintenance cost of module $i$\\
		$f_{s}({\rm a}, t)$ & & Lifetime density of the system after ${\rm a}$-th inspection\\
		$Cost ({\rm a}, \tau, i)$& & Total cost of maintenance on module $i$ at the ${\rm a}$-th inspection\\
		$EC_{sys}({\rm a},\tau)$& &Total expected cost of maintenance of the system in the interval $({\rm a}\tau, ({\rm a}+1)\tau]$ \\	
		\hline
	\end{tabular}}}\caption{Notation.}\label{tab:notation}
\end{table}

\subsection{Basic concepts}	
{PH-distributions, MAPs}, and Kronecker operations play an important role in this paper. They are the basic elements in the application of (MAMs), and are formally defined below to provide a better understanding of this work. For further details, readers are referred to (\cite{Bellman,Graham,Neuts1981}).

\begin{dfn}{{PH-distribution}}
	
	{Consider a finite Markov chain with $m$ transient
		states and one absorbing state with the infinitesimal generator $Q$ {partitioned}
		as}
	\begin{equation*}
				Q=\left( 
			\begin{array}{ c c}
				T & T^{0} \\ 
				0 & 0%
			\end{array}%
			\right).
		\end{equation*}
	
{where $T$ is a matrix of order $m$ and $T^{0}$ is a column vector
		such that $Te+T^{0}=0$. The vector $e$
		is a column vector of ones. For eventual absorption into the absorbing state,
		starting from the initial state it is necessary and sufficient that $T$
		be nonsingular. Suppose that the initial state of the Markov chain is
		chosen according to the probability vector $(\alpha,a_{m+1})$. Let 
		$X$ denote the time until absorption; then, $X$ is a random variable taking
		non-negative values, with probability distribution function given by $%
		F(x)=1-\alpha e^{Tx} e, $ for $x\geq 0.$
		
		We then denote $X$ as following a $PH(\alpha ,T)$-distribution of order $m$.}
\end{dfn}

 \begin{dfn}{Markovian Arrival Process}\\
	{Suppose that} $D=\left( d_{ij}\right) $ {is the generator of an irreducible Markov chain with} $m$ {states. At the
		end of a sojourn time in state} $i$  {that is exponentially distributed with
		parameter} $\lambda _{i}$, one of the following two events could occur: with
		probability $p_{ij}^{(1)}$, the transition corresponds to an arrival and the
		underlying Markov chain is in state $j$ with $1\leq i,j\leq m$, while with
		probability $p_{ij}^{(0)}$ the transition corresponds to no arrival and the
		state of the Markov chain is $j, j\neq i$. We can define matrices $D_{0}=\left(
		d_{ij}^{(0)}\right) $ and $D_{1}=\left( d_{ij}^{(1)}\right) $ such that  $	d_{ii}^{(0)}=-\lambda _{i},$ $d_{ij}^{(0)}=\lambda _{i}p_{ij}^{(0)},$ for $%
		j\neq i$ and $d_{ij}^{(1)}=\lambda _{i}p_{ij}^{(1)},$ $1\leq i,j\leq m$. By
		assuming $D_{0\text{ }}$ to be a nonsingular matrix, the inter-arrival times
		are finite with a probability of one, and the arrival process does not
		terminate. Hence, it can be seen that $D_{0}$ is a stable matrix. The generator $D$ is then provided by $D=D_{0}+D_{1}.$ Let $\alpha$ be the initial probability vector of the underlying Markov chain.
		
		Then, $D_{0}$ governs the transitions corresponding to no arrival and $D_{1}$ governs those corresponding to an arrival. It can be shown that MAP is
		equivalent to Neuts' versatile Markovian point process. The point process
		described by the MAP is a special class of semi-Markov processes with their transition probability matrix provided by
	\begin{equation*}
				\int_{0}^{x}e^{D_{0}t}dtD_{1}=\left[ I-e^{D_{0}x}\right] (-D_{0})^{-1}D_{1},%
			\text{ }x\geq 0
	\end{equation*}
	
	{This MAP is represented by the MAP $(D_{0},D_{1})$ of order $m$.}
\end{dfn}

\begin{dfn}{Kronecker product of matrices}\\
	If $A$\ and $B$ are rectangular matrices with dimensions $m_{1}\times m_{2}$ and
	$n_{1}\times n_{2}$, respectively, their Kronecker product $A\otimes B$ is a
	matrix with the dimensions $m_{1}n_{1}\times m_{2}n_{2}$, which can be written in compact form as
	$(a_{ij}B).$
\end{dfn}

\begin{dfn}{Kronecker sum of matrices}\\
If $A$\ and $B$ are square matrices with dimensions $m_{1}$ and $n_{1}$, respectively, their Kronecker sum, denoted by  $A\oplus B$, is a matrix defined by $A\otimes I_{n_{1}} + I_{m_{1}}\otimes B$, where $ I_{m_{1}}$, $I_{n_{1}}$ are identity matrices with dimensions $ m_{1}$ and $n_{1}$, respectively.
\end{dfn}

\section{System description}\label{sec:system}

A system formed by $N$ units grouped in $K$ modules is considered. The failure of each module can be due to the wear-out of its units or to a shock that affects the module. The lifetime of each unit follows a PH-distribution. Shocks arrive at the module (system) following a MAP. 
The generator of such system was built in \cite{MOMA2022}. All the expressions appearing in this section were derived in \cite{MOMA2022}. Here we summarize the assumptions and the results needed for the paper.

\subsection{Model assumptions} \label{sec:assumptions}
\begin{enumerate}
\item  The $i$-th module is formed by $n_{i}$ units. The lifetime of unit $j$ in the module follows a PH-distribution $PH(\alpha_{j,i},T_{j,i})$, with $m_{j,i}$ phases where $i=1,2, \cdots, K$, $j=1,2,\cdots, n_{i}$ and $N=\sum_{i=1}^{K} n_{i}$. The units within a module are considered independent. 
\item Shocks arrive to module $i$ following a $MAP(D_{0,i}, D_{1,i})$, of order $b_{i}$ where $D_{i}=D_{0,i}+D_{1,i}$ is the infinitesimal generator. Matrix $D_{0,i}$ governs the inter-arrival times between shocks that affect the module and matrix $D_{1,i}$ contains the transition rates between the phases of the MAP when a shock arrives.
\item The MAP process affecting module $i$ is independent of the other MAP processes affecting the rest of the modules.

\item Following the bottom-up strategy explained in \cite{MOMA2022}, the state of the system is ultimately determined by the states of the units that are part of each module. We consider the following notation. Let $E$ be the set of states of the system.  We denote $\widetilde{\bf s}$ one particular element of $E$, which we describe as follows

$$\widetilde{\bf s} =( \widetilde{s}_1; \widetilde{s}_2; \ldots; \widetilde{s}_K),$$

where $\widetilde{s}_i= (s_{i1}, s_{i2} , \ldots , s_{i n_i})$ denotes the state of module $i$ and is given by a particular combination of the states of the units within the module. That is $s_{ji}$ is the state of the $j$-th unit of the module $i$, with $j=1,\ldots, n_i$; and $i=1, \ldots, K$. Notice that we use ``$,$'' to discriminate states of the units within a module and ``$;$'' to distinguish the states of two different modules. Finally, let $n$ denote the total number of states of the system.

\item A shock may or may not cause the failure of a module. Let $p_{1,i}, i=1,\cdots,K$ be the probability that a shock causes the failure of module $i$, and $p_{0,i}$ the probability that the module does not fail due to the shock. Additionally, $p_{0,i}+p_{1,i}=1$.
\item The system might fail even if some modules are still operational.
\end{enumerate}

Below we provide the expression of the system generator ($\mathbb Q_{sys}$). This generator is given in terms of the number of modules that have not failed, i.e. the macro-states in the generator  refer to the number of modules that remain operational.

\begin{equation} \label{eqn:Qsys}
\mathbb Q_{sys}=
\left(
\begin{array}{c c c c c c}
{\bf Q}_{K}^{\prime} & \widetilde{\bf Q}_{K}^{\prime} & 0 & 0 & \cdots & 0\\
0 & {\bf Q}_{K-1}^{\prime} &\widetilde{\bf Q}_{K-1}^{\prime} & 0 & \cdots & 0\\
0 & 0 & \ddots & \ddots  & \cdots & 0\\
\vdots & \vdots  & &  \ddots & \ddots & \vdots \\
0 & 0 & 0 & 0 & {\bf Q}_{K-(K-1)}^{\prime}&\widetilde{\bf Q}_{K-(K-1)}^{\prime}\\
0 & 0 & 0 & 0 & 0 & 0
\end{array}
 \right).
\end{equation}

In this generator, the transitions between the macro-states are given by:

\begin{itemize}
\item ${\bf Q}_{K-l}^{\prime}$ is a matrix describing the transitions when $l$ modules out of $K$ modules have failed and the rest remain operational. Transitions in this matrix have to consider all possible combinations of $l$ modules failing out of $K$, therefore this matrix is composed of $\binom{K}{l} \times \binom{K}{l}$ blocks of matrices. The blocks in the diagonal contain the transitions within the same state i.e. from the state where $l$ specific modules have failed to the state where exactly the same modules have failed. The off-diagonal blocks in this matrix are given by matrices of zeros of appropriate dimensions, as these transitions mean that a different set of modules have failed and that is not possible. 
The diagonal blocks are akin to the following one:

\begin{equation} \label{eqn:Qi}
{\bf Q}_{1}^{s}\oplus \cdots\oplus {\bf Q}_{h_{1}-1}^{s} \oplus {\bf Q}_{h_{1}+1}^{s} \oplus \cdots \oplus {\bf Q}_{h_{l}-1}^{s}\oplus {\bf Q}_{h_{l}+1}^{s} \oplus \cdots \oplus{\bf Q}_{K}^{s}.
\end{equation}

with $h_{r}$ in $\{1,2, \cdots, K\}$, $r=1,\cdots,l$ and $\oplus$ is the Kronecker sum.

In \eqref{eqn:Qi}  matrix ${\bf Q}_{i}^{s}$ provides the transitions between the operational states in module $i$. These transitions could be due to the change of phase of its units (wear-out of the units) or to the arrival of a shock that does not affect the module. This matrix is given by,

\begin{equation*}
{\bf Q}_{i}^{s}= {\bf Q}_{i}\oplus (D_{0,i}+p_{0,i}D_{1,i}),
\end{equation*}

where matrix ${\bf Q}_{i}$ is the matrix that represents the internal changes in module $i$, i.e. changes between the operational phases of the units, that do not lead to the module failure.

\item Matrix $\widetilde{\bf Q}_{K-l}^{\prime}$ is formed by $\binom{K}{l} \times \binom{K}{l+1}$ blocks of matrices that describe the transitions when a module fails, e.g. module $p$, out of the remaining operational ones, i.e. from $K-l$ specific operational modules to $K-(l+1)$, where $l$ of the modules have previously failed. These transitions can be described by matrices like the following one:

\begin{equation}  \label{eqn:Qihat}
I_{\sum_{j=1}^{n_{1}} m_{j,1}}\otimes  \cdots  \otimes I_{\sum_{j=1}^{n_{p-1}} m_{j,p-1}}\otimes (\widetilde{\bf Q}_{p}^{s} { e}) \otimes I_{\sum_{j=1}^{n_{p+1}} m_{j,p+1}} \otimes \cdots \otimes I_{\sum_{j=1}^{n_{K}} m_{j,K}},
\end{equation}

with ${ e}$ a column vector of 1s of the appropriate dimension. The rest of the transitions  in matrix   $\widetilde{\bf Q}_{K-l}^{\prime}$ are given by matrices of zeros of appropriate dimension.

Note that in this case module $p$ fails out of the remaining operational modules, being module $p$ not necessarily consecutive to any of the previously failed ones.

In \eqref{eqn:Qihat} matrix $\widetilde{\bf Q}_{i}^{s}$ represents the transitions when the module fails due to the failure of one of its units or due to a shock. Matrix $\widetilde{\bf Q}_{i}^{s}$ is given by:

\begin{equation*}
\widetilde{\bf Q}_{i}^{s}=
\left(
\begin{array}{c|c}
\widetilde{\bf Q}_{i}\otimes I & I\otimes p_{1,i}D_{1,i}
\end{array}
\right).
\end{equation*}

where matrix $\widetilde{\bf Q}_{i}$ is the matrix that represents the transitions to a failure state due to an internal failure, i.e., the failure of its units as a consequence of the wear-out process.

\item Finally, for any $i$ in $\{1,...,K\}$, matrix ${\bf Q}_{i}^{*s}$ is the generator of module $i$ considering that it might fail due to the wear-out process or due to the arrival of a shock. This generator is given by the following expression:

\begin{equation*}
{\bf Q}_{i}^{*s}=
\left(
\begin{array}{c|c }
{\bf Q}_{i}^{s} & \widetilde{\bf Q}_{i}^{s}\\
\hline
0 & 0 \\
\end{array}
 \right).
\end{equation*}

\end{itemize}

\subsection*{Example. A $k$-out-of-$K$ system}
The macro-states of the system can be grouped into operational and failure states depending on its arrangement, for the particular case of a $k$-out-of-$K$ system, $K-k+1$ modules must fail for the system to fail. Let $\mathbb Q_{sys}^{\prime}$ be the matrix that describes the transitions between the operational states, and $\widetilde{\mathbb Q}_{sys}^{\prime}$ the matrix that describes the transitions between the operational and failure states. These matrices are given by,

\begin{equation*}
\mathbb Q_{sys}^{\prime}=
\left(
\begin{array}{c c c c c c}
{\bf Q}_{K}^{\prime} &\widetilde{\bf Q}_{K}^{\prime} & 0 &  \cdots & 0 \\
0 & {\bf Q}_{K-1}^{\prime} & \widetilde{\bf Q}_{K-1}^{\prime}  & \cdots & 0 \\
0 & 0 & \ddots & \ddots  & \cdots  \\
\vdots & \vdots  &  & \ddots & \widetilde{\bf Q}_{k+1}^{\prime} \\
0 & 0 & 0 & 0 & {\bf Q}_{k}^{\prime}\\
\end{array}
 \right)\nonumber,
\end{equation*}

and,

\begin{equation*}
\widetilde{\mathbb Q}_{sys}^{\prime}=
\left(
\begin{array}{c}
0\\
0\\
\vdots\\
0\\
\widetilde{\bf Q}_{k}^{\prime}
\end{array}
\right)\nonumber.
\end{equation*}

Therefore, the generator of the system can be described in a compact form as follows:

\begin{equation*}
\mathbb Q_{sys}=
\left(
\begin{array}{c|c}
\mathbb Q_{sys}^{\prime} & \widetilde{\mathbb Q}_{sys}^{\prime} \\
\hline
0 & 0\\
\end{array}
 \right),
\end{equation*}
As it can be seen, this generator is given in terms of 2 macro-states, the operational state (up state) that will be denoted by U and the failure state (down state) that will be denoted by D.

As for the system, we can also grouped the states of each module into operational and failure states. In \cite{MOMA2022} $Q_{i}^{*}$ is the generator that describes how the module might fail solely due to the wear-out of its units, and is given in terms of the operational units, in a similar way to the description of the failure of the system in terms of the operational modules see (\ref{eqn:Qsys}). More specifically,
	
\begin{equation*} 
{\bf Q}_{i}^{*}=
\left(
\begin{array}{c c c c c c}
Q_{n_{i}} & \widetilde{Q}_{n_{i}} & 0 & 0 & \cdots & 0\\
0 & Q_{n_{i-1}} &\widetilde{Q}_{n_{i-1}} & 0 & \cdots & 0\\
0 & 0 & \ddots & \ddots  & \cdots & 0\\
\vdots & \vdots  & &  \ddots & \ddots & \vdots \\
0 & 0 & 0 & 0 & Q_{1}&\widetilde{Q}_{1}\\
0 & 0 & 0 & 0 & 0 & 0
\end{array}
 \right),
\end{equation*}	

The macro-states of this generator can be in turn grouped into operational and failure macro-states depending on the configuration of the units within the module, e.g. if the units in the module are set in series then

\begin{center}
${\bf Q}_{i}= Q_{n_{i}}$ and $\widetilde{\bf Q}_{i}=\widetilde{Q}_{n_{i}}$.
\end{center}
where the rest of the blocks in matrix ${\bf Q}_{i}^{*}$ are given by matrices of $0s$ of appropriate dimensions.

\begin{rmk}
Matrix $\mathbb Q_{sys}$ is the generator that governs the evolution of the system until it fails. In the described system we have considered that independent MAPs can affect individual modules. Alternatively, in \cite{MOMA2022} the case where a shock that arrives following a MAP and affects the system as a whole is also considered. We do not describe this particular case in the present paper but in the next sections, matrix $\mathbb Q_{sys}$ might represent the evolution of the system standing shocks that affect the system as a whole as well.
\end{rmk}

\subsection {The {\it MoMA} algorithm}

The {\it MoMA} algorithm was developed in \cite{MOMA2022} to built the system generator using a bottom up approach, i.e. starting from the units generators. In this algorithm first the generator of module \textit{i} is built considering the units generators, then the system generator is built combining the module generators as well as the generators corresponding to the shocks arrival. Three scenarios are considered: 1. There are no shocks affecting the system, 2. A shock could affect the system as a whole causing its failure and 3. $K$ independent shock processes might affect the modules causing their failure.

\section{Maintenance policy}\label{sec:maintpolicy}

For the above described system we propose the following maintenance policy: the system is inspected at times $\tau_{\rm a}={\rm a}\tau$, where $\tau$ is fixed and determined considering maintenance costs as it will be shown later on. Additionally, maintenance actions depend on the state of the system at inspection time $\tau_{\rm a}$. For this system we distinguish between optimal operational states, critical states (the performance of the system has degraded, it could mean that although it has not failed it might be approaching the end of its life) and down states (the system has failed). The set of states that are optimal are labeled by $U_{1}$, the set of critical states are labeled by $U_{2}$ and the set of down states are labeled by $D$, although we are only considering one down state and do not distinguish failure modes.

As mentioned above, the maintenance policy looks at the state of the system at the inspection time and proceeds as follows: 
\begin{itemize}
	\item If the system is in an optimal state $\left(U_{1}\right)$, no intervention takes place, leaving the system in the same state. 
	\item If the system is in a critical state $\left(U_{2}\right)$, we look at the modules. The failed modules are replaced and for the rest we distinguish between modules in a critical state (approaching their failure) and those in an optimal state, for the former ones the failed units are maintained, for the latter ones no intervention takes place, i.e. the wear-out state of the module remains the same. 
	\item Finally, if the system has failed we replace it.
\end{itemize}

As it can be seen, we only inspect the modules if the system is in a critical state at inspection times. Otherwise, if the system is in an optimal state we do nothing, and if the system has failed we replace it following initial probability vector $\beta$. Additionally, when the system is in a critical state, maintenance actions restore the modules to an optimal state.

\begin{figure}[ht]
	\begin{center}
	\includegraphics[width=1\textwidth]{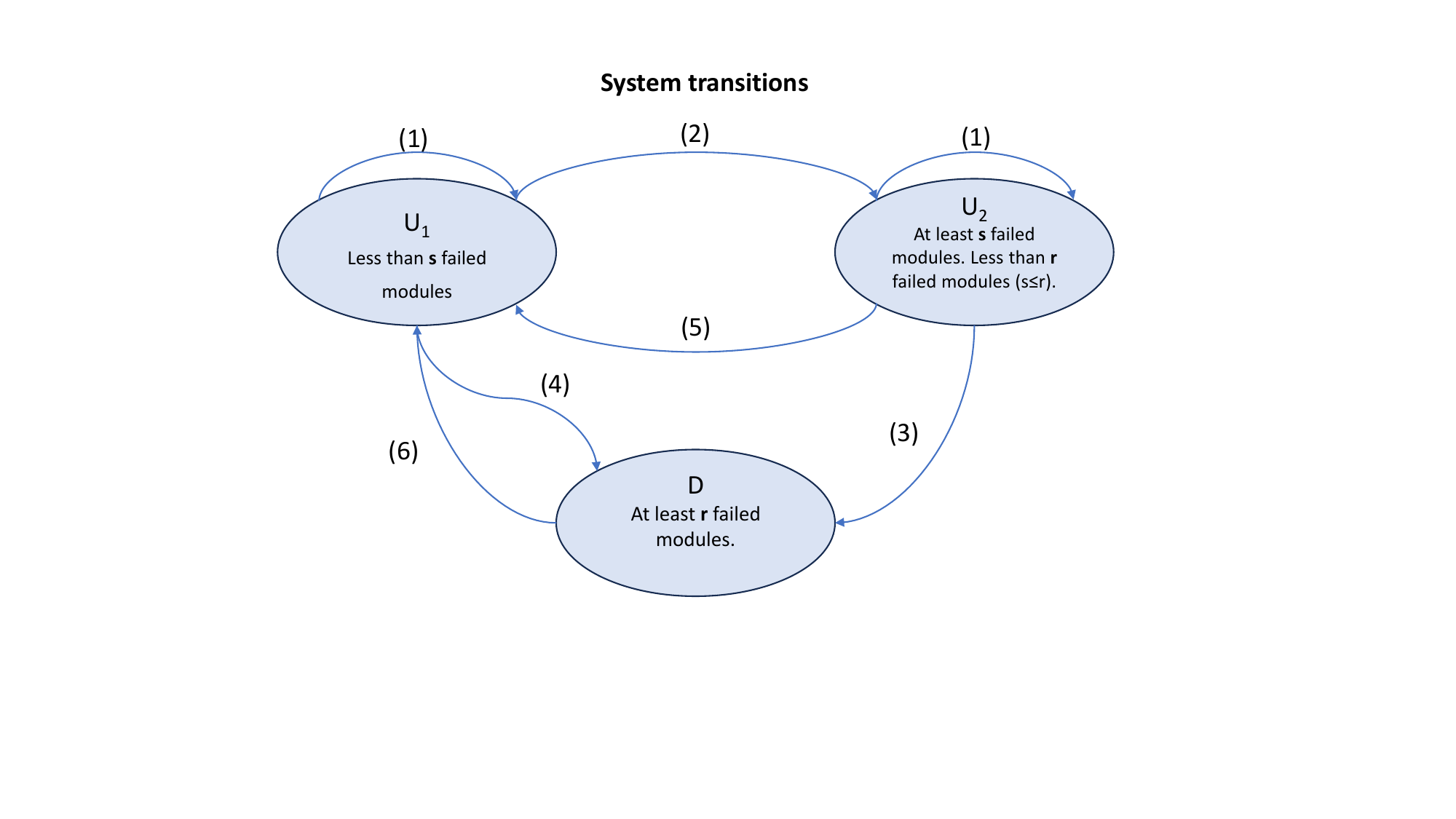}
      \vspace{-2.5cm}
	\caption{Modular system transitions. (1) Units wear-out. Shock processes. Modules might fail. (2) Failure of $s$th module. (3) Failure of $r$th module. System failure. (4) System failure due to a shock. (5) Maintenance. Restore failed units and modules. (6) Maintenance. Replace system.}	\label{fig:Transitions_graph}
\end{center}
\end{figure}

Figure \ref{fig:Transitions_graph} describes the transitions between the set of states $U_1$, $U_2$ and $D$ considering wear-out, shocks and the maintenance policy. The described transitions replicate at module level.

\subsection{Construction of maintenance matrices ${\bf M}_{i}$}
Let us define matrix ${\bf M}_{i}$ to describe how the maintenance policy affects module $i$ when the system is in a critical state. We describe this matrix by blocks. In general, let us denote $M_{A_{1}, A_{2}}$ the sub-matrix describing the transitions from states in subset $A_{1}$ to states in subset $A_{2}$, for any $A_{1}, A_{2} \subset E_{i}$ where $E_{i}=U_{i1} \cup U_{i2} \cup D_{i}$ is the state space for module $i$, for $i=1,...,K$. At inspection time $\tau_{a}$:

\begin{itemize}
\item If the module is in an optimal state, then no intervention takes place. Therefore $M_{U_{i1}, U_{i1}}=I$.
\item If the module is in a critical state, $U_{i2}$, all the failed units are maintained, the rest remain unchanged. Within $U_{i2}$ we have different combinations of units in operational and failed state. For example, let us consider a 2-out-of-4 module, this module will be in a critical state when 2 of its units have failed, we have $\binom{4}{2}$ combinations of failed and operational units that should be taken into account to determine the states within $U_{i2}$. In \ref{sec:Examples} some detailed examples are given.

In general, for a specific combination of failed units in module $i$ and after a maintenance action, module $i$ re-initiates its operation considering the following expression: 

\begin{center}
$M^{1}_{i}\otimes M^{2}_{i}\otimes \cdots \otimes M^{n_{i}}_{i}$.
\end{center}

where $M^{j}_{i}=I$ if unit $j$ is in an operational state. If the unit is in a failed state it re-initiates its operation following a maintenance action vector $\beta_{j,i}$, where $\beta_{j,i}$ contains the probabilities of the maintenance action restoring the unit to any of its operational phases.

As an example, if unit $j, j=1,\cdots,n_i$ has failed then the corresponding block of probabilities in sub-matrix $M_{U_{i1}, U_{i2}}$ is given by:

\begin{center}

$I_{m_{1,i}}\otimes  \cdots  \otimes I_{m_{j-1,i}}\otimes \beta_{j,i} \otimes I_{m_{j+1,i}} \otimes \cdots \otimes I_{m_{n_{i},i}}$.
\end{center}

\item If the module has failed, then the module is replaced and it re-initiates its operation following maintenance action vector $\beta_{i}$. Therefore, $M_{{D}, U_{i1}}=\beta_{i}$. Eventually one can take $\beta_{i}=\alpha_{i}$ (the initial probability distribution), however, to be general, we consider here that the probability of restoring module $i$ to any operational state might be different from the provided by initial probability distribution. 

\item The rest of the blocks within this matrix are given by matrices of zeros.

\end{itemize}
In summary, each row in matrix ${\bf M}_{i}$ is a vector describing the probabilities to restore the module from an specific state to any operational state after a maintenance action takes place.

\subsection{Examples}\label{sec:Examples}

Let us illustrate matrix ${\bf M}_{i}$ with a couple of examples. 

\begin{enumerate}

\item First, we consider that module $i$ is formed by 2 units set in parallel, both units are Phase-type distributed with two phases each, phases 0 and 1 are the operational phases and phase 2 is the absorption one, i.e. the unit has failed. We denote the absorption phase by F from now on. We define the operational, critical and failed states of the module in terms of the units phases as follows: $U_{i1}=\{(0,0),(0,1),(1,0),(1,1)\}$, $U_{i2}=\{(0,F),(1,F),(F,0),(F,1)\}$, and $D_{i}=\{(F,F)\}$. In this particular example the module is in a critical state when one of its units has failed. Then, matrix $M^{i}$  is given by,

\begin{center}
$
{\bf M}^{i}=
\bordermatrix{
& U_{i1} & & U_{i2} &  & D_{i}\cr
U_{i1} & I & \vrule & 0 &\vrule & 0 \cr
\hline
U_{i2} & \begin{matrix} I \otimes \beta_{2,i} \cr
 \beta_{1,i} \otimes I
\end{matrix} &  \vrule & 0 &\vrule & 0\cr
\hline
D_{i} &\begin{matrix}
\beta_{i} = \beta_{1,i}\otimes \beta_{2,i} 
\end{matrix}& \vrule & 0 &\vrule & 0}.$
\end{center}.

\item Another example is given by a module composed by 3 units, each of them Phase-type distributed with 1 phase, being 0 the operational phase and F the absorption phase, i.e. they are exponentially distributed. 

Module $i$ is a 2-out-of-3 module. Therefore the optimal states of the module are given by $U_{i1}=\{(0,0,0)\}$ and the critical states are given by $U_{i2}=\{(0,0,F),(0,F,0),(F,0,0)\}$. The down state encompasses states $\{(0,F,F),(F,0,F),(F,F,0),(F,F,F)\}$, although we do not distinguish failure modes and consider it as a unique state, as mentioned above.

The expression of matrix ${\bf M}^i$ is given by,

\begin{center}
$
{\bf M}^{i}=
\bordermatrix{
& U_{i1} & & U_{i2} & &  D_{i}\cr
U_{i1} & 1 & \vrule & 0 &\vrule & 0 \cr
\hline
U_{i2} &\begin{matrix}
 e \end{matrix} & \vrule & 0 &\vrule & 0 \cr
\hline
D_{i} & \begin{matrix}\beta_{i}=1 
\end{matrix} &  \vrule & 0 &\vrule & 0}.$
\end{center}

where $e$ is a all-ones $3\times 1$ vector.\\
\end{enumerate}

\subsection{Consequences of the maintenance actions}

Let $X(t)$ be the state of the system before the first inspection takes place. Let us recall that the system generator is given by $\mathbb Q_{sys}$ and let us say that $\alpha_{sys}$ is the system initial probability distribution.

Let us consider $\{\widetilde{X}({\rm a},t), 0\leq t \leq\tau, {\rm a}=0,1,2,\cdots,A\}$, where $\widetilde{X}({\rm a},t)$ is the state of the system at time $t\in [0,\tau]$ after the ${\rm a}$-th inspection. In other words, $\widetilde{X}({\rm a},t)$ represents the state of the system at time ${\rm a} \tau+t$. Additionally, $A$ is the maximum number of inspections, i.e at the $Ath$ inspection the system is replaced, and $\widetilde{X}(0,t)=X(t)$, $0<t\leq \tau$.

Let $E=U_{1}\cup U_{2} \cup D$. In general, the order of the states in matrix $\mathbb Q_{sys}$ allows us to directly assign the states to $U_{1}$, $U_{2}$ and $D$ without having to reorganize these states. However, this might not be always the case, when this happens we can in practice re-label the states to work with them. In what follows we assume that we do not need to reorganize nor re-label these states.

\begin{figure}[ht]
	\centering
	\includegraphics[width=1\textwidth]{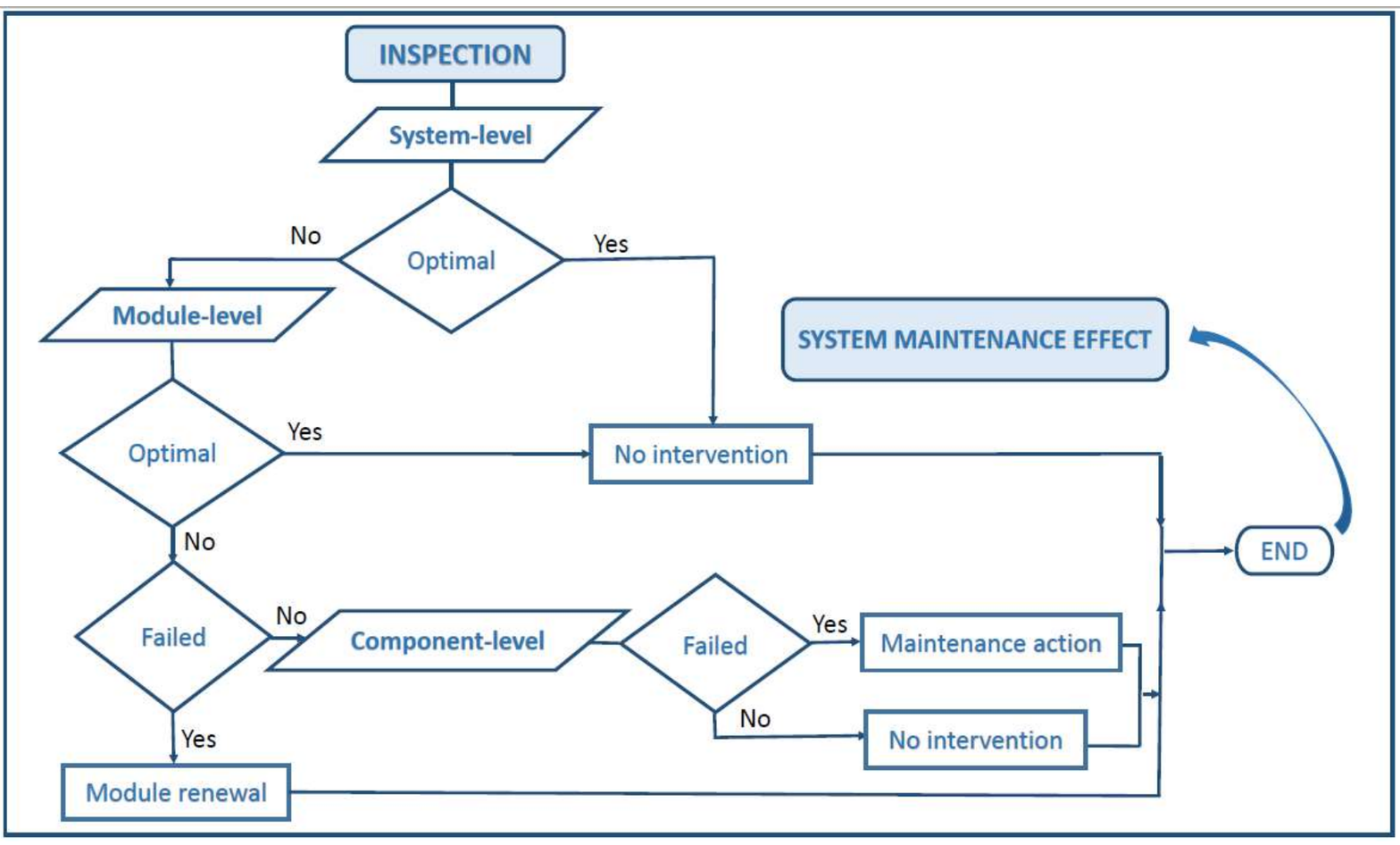}
	\caption{Maintenance scheme for the system at inspection time}	\label{fig:MSE}
\end{figure}

After the ${\rm a}$-th inspection the system is in one of the states in set $U_{1}$, i.e. $\widetilde{X}({\rm a},0) \in U_{1}$. The new state depends on the state of the system prior to the inspection and the performed maintenance action, as shown in Figure \ref{fig:MSE}, then $\widetilde{X}({\rm a},0)$ is a function of $\widetilde{X}({\rm a}-1,\tau)$ and the performed maintenance action, which is determined by matrices ${\bf M}_{1},{\bf M}_{2},...,{\bf M}_{K}$. Therefore, at the beginning of the ${\rm a}$-th period we have an initial probability distribution over the system states, let us denote this distribution $\widetilde{\alpha}({\rm a})$; if ${\rm a}=0$ then $\widetilde{\alpha}({\rm a})=\alpha_{sys}$. To determine $\widetilde{\alpha}({\rm a})$ we can distinguish three situations (see Figure \ref{fig:MSE}).

Let ${\rm a}>0$, and, for a particular state of the system $\widetilde{\bf s}\in E$, let us denote 
\[
\widetilde{\alpha}(\rm a,\widetilde{\bf s})=\PP(\widetilde{X}(\rm a,0)=\widetilde{\bf s}).
\] 
It is clear that $\widetilde{\alpha}(\rm a,\widetilde{\bf s})>0$ only for $\widetilde{\bf s}\in U_1$. Then we have:
\begin{enumerate}
	\item Denote $\widetilde{\alpha}_{U_1}(\rm a)$, the initial law at the $\rm a$-th interval given that the state of the system at the end of the previous inspection time is optimal, i.e.  $\{\widetilde{X}(\rm a -1, \tau) \in U_1\}$ . In this case, no maintenance action is performed leaving the system as it was before the inspection, consequently, for $\widetilde{\bf s}\in U_1$,
	\[
	\widetilde{\alpha}_{U_1}(\rm a,\widetilde{\bf s})=\PP(\widetilde{X}(\rm a-1,\tau)=\widetilde{\bf s}\mid \widetilde{X}(\rm a-1,\tau)\in U_1),
	\]
and $\widetilde{\alpha}_{U_1}(\rm a,\widetilde{\bf s})=0$, for $\widetilde{\bf s} \notin U_1$.

\item Denote $\widetilde{\alpha}_{U_2}(\rm a)$, the initial law at the $\rm a$-th interval given that the state of the system at the end of the previous inspection time is critical, i.e $\{\widetilde{X}(\rm a -1, \tau) \in U_2\}$.

If at the time of inspection the system is in a critical state, we look at the modules and act upon the modules to restore the system to $U_{1}$. In this sense, we can describe the probability distribution after the maintenance action in terms of the modules, more specifically, let $\widetilde{X}_{i}({\rm a},\tau)$ be the state of module $i$ at the ${\rm a}$-th inspection, then,

\begin{equation}\label {eq:alfai}
\widetilde{\alpha_{i}}({\rm a})=\widetilde{\alpha_{i}}({\rm a}-1)exp({\bf Q}_{i}^{*s}\tau){\bf M}_{i}, i=1,\cdots,K.
\nonumber
\end{equation}
being ${\bf Q}_{i}^{*s}$ the generator of module $i$, with $i=1,\cdots K$.

Considering the initial probability distribution for each module at the ${\rm a}$-th inspection we obtain the initial probability distribution of the system as follows

\begin{equation*}
\widetilde{\alpha}_{U_2}({\rm a})=\bigl(\widetilde{\alpha}_{U_{11}}({\rm a}) \otimes \widetilde{\alpha}_{U_{21}}({\rm a}) \otimes \ldots \otimes \widetilde{\alpha}_{U_{K1}}({\rm a}), 0,\cdots, 0\bigr).
\end{equation*}

given that vector $\widetilde{\alpha_{i}}({\rm a})$ can be partitioned as $(\widetilde{\alpha}_{U_{i1}}({\rm a}) \ \vrule \ \widetilde{\alpha}_{U_{i2}}({\rm a}) \ \vrule \ \widetilde{\alpha}_{D_{i}}({\rm a}))$. Notice that also in this case we have that $\widetilde{\alpha}_2(\rm a,\widetilde{\bf s})=0$, for $\widetilde{\bf s} \notin U_1$.

\item Finally, denote $\widetilde{\alpha}_{D}(\rm a)$, the initial law at the $\rm a$-th interval given that  the system has failed at the end of the previous inspection time, i.e. $\{\widetilde{X}(\rm a -1, \tau) \in D\}$. In this case, the system it is replaced following maintenance action vector $\beta$, then $\widetilde{\alpha}_3(\rm a)=\beta$.

\end{enumerate}
\medskip

\noindent So, the initial probability distribution for the system at inspection time $\tau_{\rm a}={\rm a}\tau$ taking into account these situations is given by

\begin{eqnarray}\label {eq:alfa}
\nonumber &&\widetilde{\alpha}({\rm a})=\widetilde{\alpha}_{U_1}({\rm a})\PP(\widetilde{X}({\rm a-1},\tau) \in U_{1}) \\
\nonumber &&\hspace{1.25cm}  +\widetilde{\alpha}_{U_2}({\rm a})\PP(\widetilde{X}({\rm a-1},\tau) \in U_{2}) \\
&&\hspace{1.25cm} + \widetilde{\alpha}_{D}(\rm a)\PP(\widetilde{X}({\rm a-1},\tau) \in D).
\end{eqnarray}

\section{Maintenance cost}\label{sec:cost}
Every maintenance action has an associated cost that comprises the inspection cost, the cost associated to the performed maintenance action, the set-up cost, etc. Additionally, the maintenance action cost depends on the performed maintenance action and this in turn depends on the state of the system at the inspection time $\tau_{{\rm a}}$.

\subsection*{Case 1: The system is in an optimal state} 

If the system is in an optimal state, no maintenance action takes place, therefore, we just have an inspection cost given by $C_{I}$. 

\subsection*{Case 2: The system is in a critical state}

In this situation we define matrix ${\bf C}_{i}$ the cost matrix associated to the maintenance actions performed to each module at the ${\rm a}$-th inspection. This matrix is described by blocks in a similar way to matrix ${\bf M}_{i}$:

\begin{itemize}

\item If the module is in an optimal state, then no intervention takes place and we just have the inspection cost. Therefore $C_{U_{i1}, U_{i1}}=C_{I}I$, where $I$ is an identity matrix with the same dimensions as $U_{i1}$.
\item If the module is in a critical state, all the failed units are maintained, the rest remain unchanged. We have an associated cost for each of the failed units, that depends on the unit state and also on the performed maintenance action. 

Within block $C_{U_{i2}, U_{i1}}$ we have different combinations of units in operational and failed state. For a specific combination we have a cost associated to the maintenance action performed on the failed units, i.e. the total cost to restore the module to an optimal state would be given by the sum of the costs associated to each of the failed units.

In practice, to build this block in matrix ${\bf C}_{i}$, we consider a cost vector that includes the cost associated to restore each unit to one of the optimal phases plus the inspection cost. Then, for each specific combination of failed and operational units we multiply the cost vector for a vector that contains 1 if the unit has failed and the transition to the optimal phase is possible, and 0 otherwise. For example, let us consider a module with 3 units set in parallel, the first unit is of Phase-type with 2 phases (0, 1 and F being the absorption phase), and the second and third units are of Phase-type with one phase (0 and F being the absorption phase). Let us consider the situation where the first and third unit have failed, we want to calculate the cost of maintaining this module in a critical state. We have two possibilities: 

\begin{enumerate}
\item We could go from state (F,0,F) to state (0,0,0), in this case the cost would be given by $c_{F0}^1+c_{F0}^3=(1,0,0,1)(c_{F0}^1, c_{F1}^1, c_{F0}^2, c_{F0}^3)^{t}$ where $c_{F0}^1$ is the cost associated to restoring unit 1 from failure to phase 0 plus the inspection cost.

\item We could go from state (F,0,F) to state (1,0,0), in this situation the cost would be given by $c_{F1}^1+c_{F0}^3=(0,1,0,1)(c_{F0}^1, c_{F1}^1, c_{F0}^2, c_{F0}^3)^{t}$ where $c_{F1}^1$ is the cost associated to restoring unit 1 from failure to phase 1 in addition to the inspection cost.
\end {enumerate}

\item If the module has failed then it is replaced with a cost replacement vector $C_{RM,i}$. This vector contains the cost of re-initializing the module to any of its optimal phases after the maintenance intervention plus the inspection cost. See the application in Section \ref{sec:application}.

\subsection*{Example. 2-units parallel module}
To illustrate matrix ${\bf C}_{i}$ we consider a previous example of a module formed by 2 units set in parallel, being both units Phase-type distributed with two phases. For this particular example, the cost associated to maintaining the units is given by the cost vector $c_{i}=(c_{F0}^1, c_{F1}^1, c_{F0}^2, c_{F1}^2)^{t}$. We have that matrix ${\bf C}_{i}$ is given by \\

${\bf C}_{i}=$
\[
\bordermatrix{
& U_{i1} & & U_{i2} &  & D_{i}\cr
U_{i1} & C_{I}I & \vrule & 0 &\vrule & 0 \cr
\hline
U_{i2} & \begin{matrix} (0,0,1,0) c_{i} &  (0,0,0,1) c_{i} & 0 & 0 \cr
0 & 0  & (0,0,1,0) c_{i} & (0,0,0,1) c_{i} \cr
(1,0,0,0) c_{i} &  0 & (0,1,0,0) c_{i} & 0 \cr
0 & (1,0,0,0) c_{i}  & 0 & (0,1,0 , 0) c_{i} \cr
\end{matrix} &  \vrule & 0 &\vrule & 0\cr
\hline
D_{i} &\begin{matrix}
C_{RM,i} 
\end{matrix}& \vrule & 0 &\vrule & 0}
\]

equivalently,

\[
\bordermatrix{
& U_{i1} & & U_{i2} &  & D_{i}\cr
U_{i1} & C_{I}I & \vrule & 0 &\vrule & 0 \cr
\hline
U_{i2} & \begin{matrix}  c_{F0}^2 &   c_{F1}^2 & 0 & 0 \cr
0 & 0  & c_{F0}^2 &   c_{F1}^2 \cr
c_{F0}^1 &  0 & c_{F1}^1  & 0 \cr
0 & c_{F0}^1 &  0 &  c_{F1}^1\cr
\end{matrix} &  \vrule & 0 &\vrule & 0\cr
\hline
D_{i} &\begin{matrix}
C_{RM,i}
\end{matrix}& \vrule & 0 &\vrule & 0}
\]
\end{itemize}

\subsection*{Case 3: The system has failed}

Finally, if the system is in a failed state and needs to be replaced, the cost associated is given by the inspection cost $C_I$, the replacement cost $C_{SR}$ and the loss of production cost due to the the system being down during certain interval of time. For the latter we calculate the expected loss of production cost, or down cost. Let us define $c_{down}({\rm a}, t)$ a function that provides the cost due to the system downtime at the ${\rm a}$-th interval $({\rm a}\tau,({\rm a}+1)\tau]$. If the system has failed at time ${\rm a}\tau+t$, where $0<t<\tau$, then the expected down cost in the interval $(0,\tau]$ is given by:

\begin{equation}\label{eq:cost_down}
EC_{down}({\rm a},\tau)=\int_{0}^{\tau} f_{s}({\rm a}, t)c_{down}({\rm a},t) {\rm d} t.
\end{equation}

where $f_{s}({\rm a}, t)=\widetilde{\alpha}({\rm a})exp(Q_{sys}t)(-Q_{sys}\textbf{e})$ is the density function of the lifetime of the system. 

As a particular case we can consider $c_{down}({\rm a}, t)$ a linear function of the non-operational time since the failure of the system until $\tau$, then $c_{down}({\rm a},t)=(\tau-t)C_{down}$, where $C_{down}$ is a constant.

Considering all the possibilities, we obtain the expected maintenance cost at the $\rm a$-th inspection time. Firstly, the cost incurred due to the maintenance of module $i$, if the system is in a critical state at the ${\rm a}$-th time, is determined as follows  
\begin{equation}\label{eq:Costi}
Cost ({\rm a}, \tau, i)=\widetilde{\alpha}_i({\rm a}-1) exp({\bf Q}_{i}^{*s}\tau){\bf D_{i,U_2}}({\bf M}_{i}\odot {\bf C}_{i})e, i=1,\cdots,K. 
\end{equation}

where $\odot$ represents the Hadamard product or element wise product. Besides, for the system to be in a critical state, only certain states of the module $i$ are allowed for the expression \eqref{eq:Costi}. More specifically, for $i=1, \ldots, K$,  let $(\widetilde{\bf s}; \widetilde{s}_i)$ denote a state of the system in which the $i$-th module is in state $\widetilde{s}_i$, regardless the rest of the modules of the system. We denote ${\bf D_{i,U_2}}$ a diagonal matrix of appropriate dimension, which has 1 in the diagonal only in those positions corresponding to states $\widetilde{s}_i$ for which there exists a combination of states of the rest of modules such that $(\widetilde{\bf s}; \widetilde{s}_i) \in U_2$. That is, to calculate the cost incurred by the maintenance action of module $i$ in this case, we only consider when module $i$ is in one of those states that lead to a critical state of the system. Therefore, we need to consider the combination of the modules states to determine the system state.

Then, the expected maintenance cost for the system is given, at inspection ${\rm a}$-th, by

\begin{eqnarray*}
EC_{sys}({\rm a},\tau)&=&C_{I}\PP(\widetilde{X}({\rm a}-1,\tau) \in U_{1})\\
&&+\left(\sum_{i=1}^{K}Cost ({\rm a}, \tau, i)\right)
\PP(\widetilde{X}({\rm a}-1,\tau) \in U_{2})  \\
&&+ \left(C_I+EC_{down}({\rm a}-1,\tau) +C_{SR}\right)\PP(\widetilde{X}({\rm a}-1,\tau) \in D).
\end{eqnarray*}

Finally, assuming a maximum number of $A$ inspections to perform along the useful life of the system, the total maintenance cost for that period is given by 
\[
EC_{sys}(\tau)=\sum_{a=1}^{A}EC_{sys}({\rm a},\tau) .
\]
Thus, an optimization problem can be formulated as a function of the incurred cost and the frequency of inspection $\tau$. We will address this problem numerically in the next section when the model is illustrated by means of a real case study.


\section{Numerical application}\label{sec:application}

Our numerical example is inspired by the case study presented in Liu, Liu and Cai (2014) (\cite{Liu2014}). The authors carry out a reliability analysis of the electrical control system of a subsea Blowout Preventer (BOP) stack based on Markov methodology. Specifically, two voting schemes on system performance are evaluated based on Markov modeling. The electrical control system of subsea BOP system is a redundant system.  As it is described in \cite{Liu2014}, two control pods are above the BOP stack, affording redundant control of subsea functions. Each pod contains a Subsea Electrical Module (SEM), which is contained in domed containers made of thick steel under external water pressure.

The SEMs contain processor modules, input modules, and output modules. Other secondary components of the system are considered completely reliable.  Three processors are used to form a processor subsystem (module). Each processor runs the same application programs, processing data and sending command signals. The processor subsystem performs 2-out-of-3 voting, which means the system can hence detect and correct a single failure. 
For the sake of simplicity, in this paper we assume that the input and output modules also performs 2-out-of-3 voting. In the surface, 2 control panels are installed  working in parallel. If control panels fail, they will be repaired immediately without removing the BOP system from the water. The subsea components namely input modules, output modules and processors will be pulled to the surface and repaired when the control system cannot work due to their failure.

We consider the same values of transition rates of \cite{Liu2014} for processors, input and output modules. That is, the lifetime of each of the 3 processors in the corresponding module has Exponential distribution with rate $\lambda_p$ as specified in Table \ref{tab:rates}. We treat the input and output modules similarly. However, for the control panels, we consider a more general case. Specifically we assume that each follows a Phase-type distribution with the parametrization given in Table \ref{tab:rates}. In this case we have chosen the parameters of the Phase-type distribution with the same mean time as the Exponential distribution considered in \cite{Liu2014} for the control panels.

\begin{table}[ht]
	\centering
	\begin{tabular}{|c|c|l|}\hline
		\textbf{Module} & \textbf{Distribution } & {\bf Parameters}\\ \hline
		Processor     & Exponential  & $\lambda_p= 1.820 $ \\ \hline
		Input         & Exponential & $\lambda_{in}= .9798$ \\ \hline
		Output        & Exponential & $\lambda_{out}= .9780$ \\ \hline
		Control Panel & Phase type  &  ${\bf Q_{c}}=\left(\begin{array}{cc}
			-6.304 &  6.304 \\
			0     & -6.304
		\end{array}\right); \ \alpha_c=(1,0)$\\ \hline
	\end{tabular}
	\caption{Lifetime distribution of the units per module. Time is scaled by 1e-5 time units. }\label{tab:rates}
\end{table}

%

%
%

\subsection{Description of the modules}
\begin{figure}[ht]
	\centering
	
	\includegraphics[width=0.9\textwidth]{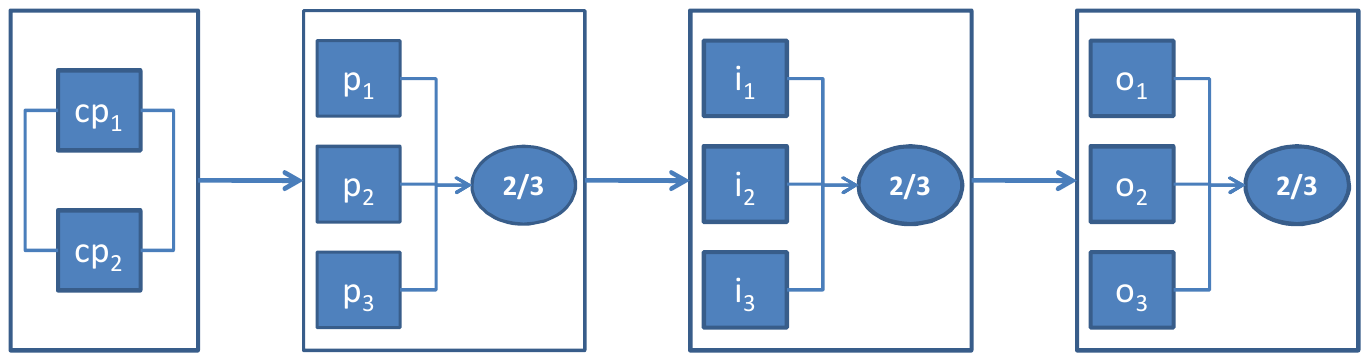}
	\vspace{-3cm}
	\caption{Reliability Block Diagram of the SEM system}	\label{fig:rbd}
\end{figure}

Following \cite{Liu2014} the SEM system consists of the 4 modules described above (control panel, processors, input and output) arranged in series. Figure \ref{fig:rbd} shows the corresponding reliability block diagram.

Let $X(t)$ denote the state of the system (SEM) at time $t$. We then have that $\{X(t), t>0\}$ is a continuous-time Markov chain and we can use the algorithm {\it MoMA} described in \cite{Gamiz2022} to derive the system generating matrix. We first analyze each module separately. 

In this system we have two types of modules, for each module we obtain its generator. In generator ${\bf Q}_{i}$ we only provide the transitions between optimal and critical states, given that the transitions to the down state can be deduced from them.
\begin{itemize}
	\item $M_1$: {\it Control panel.} This module is a two-units parallel structure. Each unit has Phase-type distribution lifetime with parameters $(\alpha_c,{\bf Q}_c)$ given in Table \ref{tab:rates}. 
	The state of $M_1$ along time is described by a Markov process $\{X_{1}(t), t>0\}$. At any time $t >0$ each unit of the module can be in one of three states: 0, if the unit is in optimal state; 1, if the unit is operative but critical; and, $F$, if the unit is in failure. Then after applying {\it MoMA} we get the set of all possible states of the module, that is, $E_1=\{(0,0),(0,1),(1,0),(1,1),(0,F),(1,F),(F,0),(F,1),(F,F)\}$. Given the parallel structure of this module we can split the state space into a subset of optimal states $U_{11}=\{(0,0),(0,1),(1,0),(1,1)\}$; a subset of critical states $U_{12}=\{(0,F),(1,F),(F,0),(F,1)\}$, and a unitary subset with the failure state $D_1=\{(F,F)\}$. The lifetime of the module has Phase-type distribution with parameters given as 
	\begin{eqnarray*}\label{eq:QM1}
		&&\hspace{-1cm}{\bf Q}_{1}=\\
		&&\hspace{-1.5cm}\left(
		\nonumber	\begin{array}{rrrrrrrr}
			-12.608 & 6.304 & 6.304 & 0.000 & 0.000 & 0.000 & 0.000 & 0.000 \\ 
			0.000 & -12.608 & 0.000 & 6.304 & 6.304 & 0.000 & 0.000 & 0.000 \\ 
			0.000 & 0.000 & -12.608 & 6.304 & 0.000 & 0.000 & 6.304 & 0.000 \\ 
			0.000 & 0.000 & 0.000 & -12.608 & 0.000 & 6.304 & 0.000 & 6.304 \\ 
			0.000 & 0.000 & 0.000 & 0.000 & -6.304 & 6.304 & 0.000 & 0.000 \\ 
			0.000 & 0.000 & 0.000 & 0.000 & 0.000 & -6.304 & 0.000 & 0.000 \\ 
			0.000 & 0.000 & 0.000 & 0.000 & 0.000 & 0.000 & -6.304 & 6.304 \\ 
			0.000 & 0.000 & 0.000 & 0.000 & 0.000 & 0.000 & 0.000 & -6.304
		\end{array}\right),
	\end{eqnarray*}
	and initial vector $\alpha_{1}=(1,0,0,0,0,0,0,0)$. 
	
	\item $M_2$: {\it Processor.} This module  is a 2-out-of-3 structure. It is composed of 3 units that are independent and identically distributed (i.i.d.) with Exponential lifetime with rate $\lambda_p= 1.820$. In this case the state space is simpler than $M_1$, that is, only two states are considered for each unit: 0, if the unit is operative, and $F$ if the unit is in failure. Let $\{X_2(t), t>0\}$ be the Markov chain describing at any time the module state. The  set of states can be partitioned into $E_2=U_{21} \cup U_{22} \cup D_2$, where  $U_{21}=\{(0,0,0)\}$ is the optimal state; $U_{22}=\{(0,0,F),(0,F,0),(F,0,0)\}$ are the critical states; and, $D_2=\{(0,F,F),(F,0,F),(F,F,0),(F,F,F)\}$ are the failure states. We remind that we do not consider transitions between the failure states so we actually treat the subset $D_2$ as a unitary set. Thus, the model considered has only 5 states in total. Using the {\it MoMA} algorithm, we obtain the module lifetime is Phase-type with parameters given by
	
	\begin{equation*}\label{eq:QM2}
		{\bf Q}_{2}=\left(
		\begin{array}{rrrr}	
			-5.46&     1.82&     1.82&     1.82\\
			0.00 &   -3.64  &   0.00  &   0.00 \\
			0.00 &    0.00 &  -3.64  &   0.00 \\
			0.00  &   0.00  &   0.00  &  -3.64 \\
		\end{array}\right), 
	\end{equation*}
	and the initial vector is given by $\alpha_2=(1,0,0,0)$.
	
	\item  $M_3$: {\it Input.} This module  is  a 2-out-of-3 structure. The units have i.i.d. exponential lifetimes with rate  $\lambda_{in}= .9798$. Let $\{X_3(t), t>0\}$ be the Markov chain describing the state of the module. The set of states of the module is equal to $M_2$ and we can consider a similar partition. That is, $E_3=U_{31} \cup U_{32} \cup D_3$, with $U_{31}$ the optimal state; $U_{32}$, the critical states, and $D_3$ the failure state of the module. The lifetime of $M_3$ has Phase-type distribution with parameters 
	\begin{equation*}\label{eq:QM3}
		{\bf Q}_{3}=\left(
		\begin{array}{rrrr}	
			-2.939 & 0.980 & 0.980 & 0.980\\ 
			0.000 & -1.960 & 0.000 & 0.000  \\ 
			0.000 & 0.000 & -1.960 & 0.000  \\ 
			0.000 & 0.000 & 0.000 & -1.960 \\ 
		\end{array}\right), 
	\end{equation*}
	and  $\alpha_3=(1,0,0,0,0)$.
	
	\item  $M_4$: {\it Output.} This module  is  a 2-out-of-3 structure. The units have i.i.d. Exponential lifetimes with rate $\lambda_{out}= .9780$. Let $\{X_4(t), t>0\}$ be the Markov chain describing the state of the module. The set of states of the module is equal to $M_2$ and we can consider a similar partition. That is, $E_4=U_{41} \cup U_{42} \cup D_4$, with $U_{41}$ the optimal state; $U_{42}$, the critical states, and $D_4$ the failure state of the module. The lifetime of $M_4$ has Phase-type distribution with parameters 
	\begin{equation*}\label{eq:QM4}
		{\bf Q}_{4}=\left(
		\begin{array}{rrrr}	
			-2.934 & 0.978 & 0.978 & 0.978 \\ 
			0.000 & -1.956 & 0.000 & 0.000 \\ 
			0.000 & 0.000 & -1.956 & 0.000 \\ 
			0.000 & 0.000 & 0.000 & -1.956 \\
		\end{array}\right), 
	\end{equation*}
	and the initial vector is given by $\alpha_4=(1,0,0,0)$.
	
\end{itemize}

\subsection{System model without maintenance}
Let $\{X(t); t>0\}$ be the Markov chain describing the system behaviour through time.  A graphical description of the system is given in Figure \ref{fig:rbd}.
The {\it MoMA} algorithm has been run again to assemble $M_1$, $M_2$, $M_3$ and $M_4$ in a series structure to compose the system. We obtain the state of the system as a function of the states of the modules, i.e. $X(t)=\Phi(X_1(t),X_2(t),X_3(t),X_4(t))$, with $\Phi$ defined in terms of the series structure of the system.

We denote the state space $E=U \cup D$, partitioned in two subsets, which are up states  and down states. After running {\it MoMA}, we obtain 512 operative states as the size of set $U$, and 133 down states. Similar to the module cases, we do not take into account transitions between the down states, thus we consider the set of down states $D$ as a unitary set. 
The system has Phase-type distribution with parameters $({\alpha}_{sys},{\mathbb Q}_{sys})$, which we do not specifically write here due to space limitations.
Moreover, we assume that at time $t=0$ all the units in the system are in optimal conditions so, all components of vector ${\alpha}_{sys}$ are equal to 0 except the first one, which is equal to 1.

Similar to the modules we distinguish optimal states and critical states in the system. That is $U= U_1 \cup U_2$, where in particular $U_1=\{(0.0:0.0.0:0.0.0:0.0.0), (0.1:0.0.0:0.0.0:0.0.0),
(1.0:0.0.0:0.0.0:0.0.0), (1.1:0.0.0:0.0.0:0.0.0)\}$ and the rest of operative states are critical. We denote a generic state $\widetilde{i}$ of the system as

$$\widetilde{s}=(s_{11},s_{12};s_{21},s_{22},s_{23};s_{31},s_{32},s_{33};s_{41},s_{42},s_{43})$$

where $s_{ji}$ denotes the state of the $j$-th unit in module $i$ being $i=1,2,3,4$ and $j=1,2,3$.

\begin{figure}[ht]
	\centering
	\includegraphics[width=0.7\textwidth]{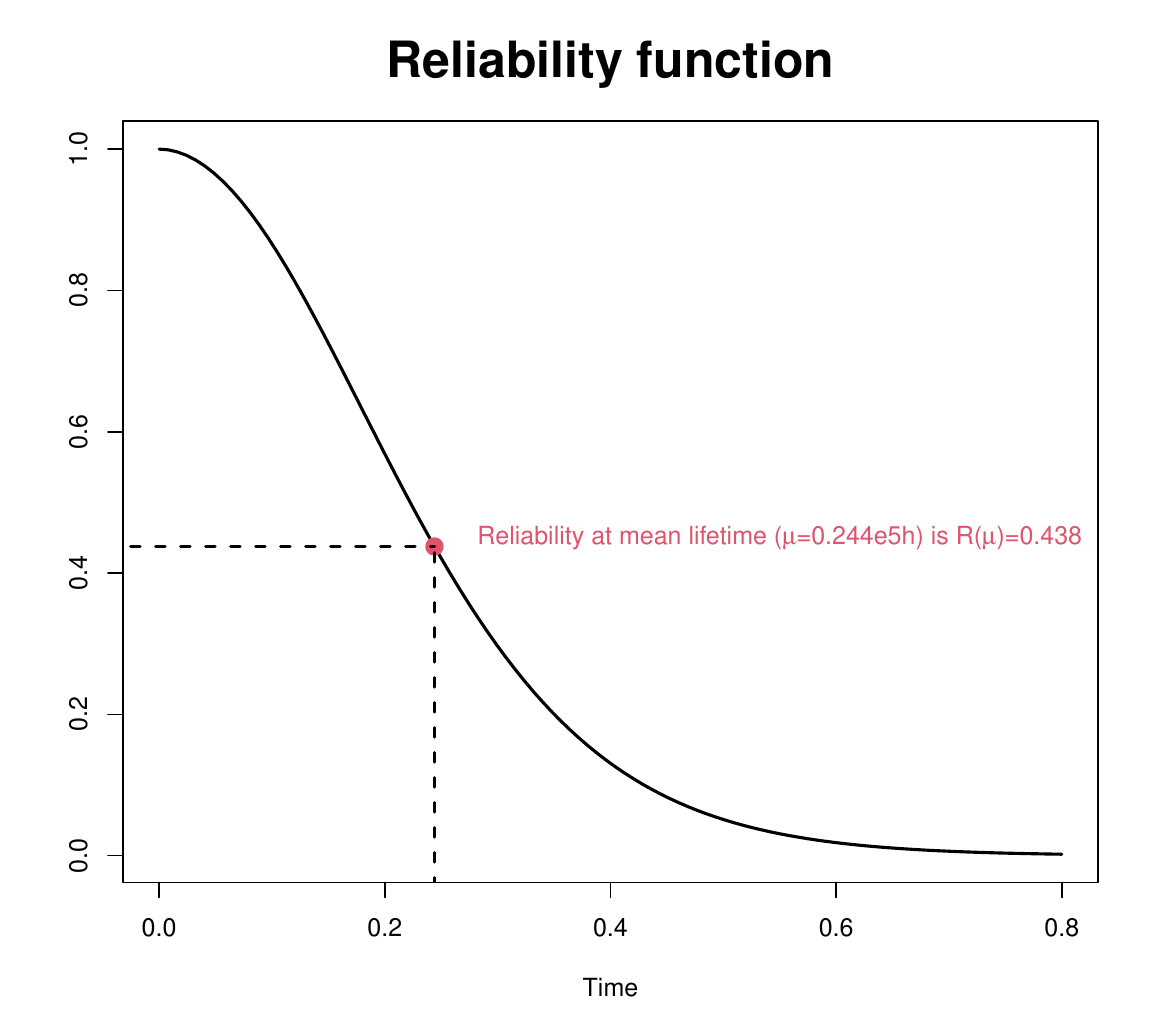}
	\caption{Reliability function of SEM system}	\label{fig:reliab}
\end{figure}

The mean lifetime of the system is $\mu=24000 \ hours$.  The reliability function is represented in Figure \ref{fig:reliab}.

\subsection{Maintenance}
Maintenance policy is explained in Section \ref{sec:maintpolicy}.  The system is assumed to be inspected regularly at times $\tau_{\rm a}= {\rm a} \tau$, with ${\rm a}=0, 1, \ldots, A$ and $\tau >0$. For ${\rm a}=0$ that is $\tau_{0}=0$ we assume that the system is operating in optimal conditions.

To describe the state of the system (with maintenance) we introduced the stochastic process $\{\widetilde{X}({\rm a},t)\}$ in Section \ref{sec:maintpolicy} , where $\widetilde{X}(\rm a,t)$ gives the state of the system at time $t+{\rm a}\tau$, with $0 \leq t \leq \tau$. 
Let us call $({\rm a} \tau, ({\rm a}+1)\tau]$ the ${\rm a}$-th \textit{operating cycle} of the system ($\rm a$-th interval).



Focusing on the first inspection period, the concrete intervention carried out depends on the state of the system at time $\tau$, and in any case, consists of restoring the system to a desired performance level.   The new state for the system is chosen according to a pre-specified maintenance rule (see Section \ref{sec:maintpolicy}). After maintenance a new operating cycle starts, and the state of the system is described by $\widetilde{X}(1,t)$, with $0 < t \leq \tau$. 
This policy is used at the end of every $\tau$-cycle.
Each maintenance action involves a particular cost that we specify in Table \ref{tab:cost} for our particular SEM system. 

\begin{table}[ht]
	\begin{center}
		\begin{tabular}{|c|c|c|}\hline
			&\textbf{Value}        & \textbf{Description}\\ \hline
			$\boldsymbol{\beta}_{j,i}$ &$(0.5,0.5)^t$ &  Vector of probabilities of restoring \\
			&& failed unit $j$ of module $i$ to operational phase \\ \hline	
			$C_{I}$              & 1       &	Cost of inspection (${\tt mu}$)\\ \hline
			$c_{F0} $            & 1         &Cost of restoring a unit to its optimal  \\
			&&	operational phase (${\tt mu}$) \\ \hline
			$C_{F1} $            & 0.5       &Cost of restoring a unit to a  \\
			&&	non-optimal operational phase (${\tt mu}$) \\ \hline
			$C_{RM}$             & $3 C_{F0}$     & Cost of replacement of a module  (${\tt mu}$) \\ \hline
			$C_{SR}$             &$3 C_{RM}$    & Cost of replacement of the system  (${\tt mu}$)  \\ \hline
			& Sc. 1: 1e-3       &  \\
			$C_{down}$    & Sc. 2: 1e-2 &Cost of non productivity \\
			& Sc. 3: 1e-1& per unit time $({\tt mu}/{\tt h})$\\
			& Sc. 4: 1 \hspace{0.4cm} \\
			\hline 	
		\end{tabular}
	\end{center}	
	\caption{Specification of values for parameters of the model for the simulated SEM system, considering several scenarii, given by 4 different values of parameter $C_{down}$. Note: ${\tt mu}$=monetary unit. }
	\label{tab:cost}
\end{table}

In this particular case we assume that the replacement costs of the different modules are the same, as well as the costs associated to restoring each unit to an operational phase. We consider 5 different cases for the non-operational cost, $C_{down}$. See Section \ref{sec:maintpolicy}.

The initial state in a new operating cycle, i.e. $\widetilde{X}(1,0)$ is selected according to a probability vector $\widetilde{\alpha}(1)$ which is defined according to equation \eqref{eq:alfa}. We provide here a reminder. 

\subsection*{Case 1. The system is in optimal conditions} If $\widetilde{X}(0,\tau) \in U_1$, the system is not maintained, and then $$\PP(\widetilde{X}(1,0)=\widetilde{s})=\PP(\widetilde{X}(0,\tau))=\widetilde{s})=\PP(X(\tau)=\widetilde{s}).$$ where $\widetilde{s}$ has been defined above.
That is, the initial vector for the second operating cycle $\widetilde{\alpha}(1)$ is given by the occupation probability vector at time $\tau$. In this case the only cost involved is the inspection cost $(C_{I})$. \\

\subsection*{Case 2. The system is in failure} If $\widetilde{X}(0,\tau) \in D$, the full system is replaced. Then $\widetilde{\alpha}(1)=\beta$. In this case the involved cost is, on the one hand $C_{SR}$ plus the expected cost due to non-productivity cost given in equation \eqref{eq:cost_down}.

\subsection*{Case 3. The system is critical} If $\widetilde{X}(0,\tau) \in U_2$, the system is maintained. As explained in Section \ref{sec:maintpolicy}, every module of the system is inspected separately; and depending on the state of the module an action is undertaken, as explained below. For $i=1,2,3,4$, let us denote $\{\widetilde{X}_i({\rm a},t); {\rm a}=0,1,\cdots, A; 0 < t \leq \tau\}$ the process that represents the state of module $M_i$ when maintenance is considered. 
In this case to determine the initial probability law we consider the state selected for each module after being maintained, $\widetilde{\alpha}_i(1)$, for $i=1,2,3,4$ and that will be  the starting point of the module for the second $\tau$-period. The initial vector of probabilities at the system level is then 
$$\widetilde{\alpha}(1) =\widetilde{\alpha}_1(1)\otimes \widetilde{\alpha}_2(1)\otimes\widetilde{\alpha}_3(1)\otimes \widetilde{\alpha}_4(1).$$

Note that we only focus here on the transitions to the optimal set of states $U_{1}$ given that the rest of the transitions are not possible.

\begin{itemize}
	\item {\textbf{Module} $M_1$.} This is a 2-unit parallel module. The units are Phase-type with 2 operative phases.
	\begin{itemize}
		\item  The module state is critical if one unit has failed $(U_{12})$. This unit is then restored to one of the two operative phases according to a vector of probabilities $\beta_{j,1}$, $j=1,2$. The operative unit remains unchanged. 
		
		\item The module has failed if both units have failed. In this case, the module is restored to the optimal state class $U_{11}$ according to the vector of probabilities $\beta_{1,1}\otimes \beta_{2,1}$. 
	\end{itemize}
	In the particular case of the example we take $\beta_1=\beta_2=(0.5,0.5)$, then the maintenance action on this module is ruled by matrix ${\bf  M}_1$ given in \eqref{eq:mainM1}. \\
	
	Each maintenance action involves a particular cost determined by the corresponding transition of the matrix ${\bf M}_1$. If the module state is such that $\widetilde{X}_1(0,\tau) \in U_{11}$, the only cost is due to inspection. If $\widetilde{X}_1(0,\tau) \in U_{12}$, then the failed units are maintained and we have to consider the cost of restoring these units to an operational state in addition the inspection cost. If $\widetilde{X}_1(0,\tau) \in D_{1}$, then the module is restored to $U_{11}$, the incurred cost here is the inspection cost plus the renewal cost which depends on the state of $U_{11}$ that is eventually chosen. 
	
	The matrices of maintenance action ${\bf M}_1$ and incurred cost ${\bf C}_1$  are obtained as
	
	\begin{equation}\label{eq:mainM1}
		\hspace{-2.5cm}	{\bf M}_1=\left(
		\begin{array}{cccc}
			1.00 & 0.00 & 0.00 & 0.00 \\ 
			0.00 & 1.00 & 0.00 & 0.00 \\ 
			0.00 & 0.00 & 1.00 & 0.00 \\ 
			0.00 & 0.00 & 0.00 & 1.00 \\ 
			0.50 & 0.50 & 0.00 & 0.00 \\ 
			0.00 & 0.00 & 0.50 & 0.50 \\ 
			0.50 & 0.00 & 0.50 & 0.00 \\ 
			0.00 & 0.50 & 0.00 & 0.50 \\ 
			0.25 & 0.25 & 0.25 & 0.25 \\ 
		\end{array}
		\right), \hspace{0.3cm}
		{\bf C}_1=\left(
		\begin{array}{cccc}
			1 & 0 & 0 & 0 \\ 
			0 & 1 & 0 & 0 \\ 
			0 & 0 & 0 & 0 \\ 
			0 & 0 & 0 & 1 \\ 
			3 & 4.5 & 0 & 0 \\ 
			0 & 0 & 3 & 2.5 \\ 
			3 & 0 & 2.5 & 0 \\ 
			0 & 3 & 0 & 2.5 \\ 
			3 & 2.5 & 2.5 & 2 \\ 
		\end{array}
		\right)
	\end{equation}
	In addition to the costs due to unit repair, unit replacement or module replacement (whichever intervention has been carried out) the inspection cost has to be taken into account.

	\item {\bf Modules} $M_2$, $M_3$, $M_4$. The rest of modules in the SEM are similar, each consisting of 2-out-of-3 of exponentially distributed units, in all cases the optimal set of states is a unitary set where all three units in the module are functional. In all cases, the maintenance matrix reduces to a column vector of 1; and the corresponding cost matrix also reduces to a column vector such that 
	${\bf c}_{i}=(C_{I},C_{I}+c_{F0}^{1}, C_{I}+c_{F0}^{2},C_{I}+ c_{F0}^{3}, C_{I}+C_{RM,i})^t$. Considering the values in Table \ref{tab:cost}, we have  ${\bf c}_{i}=(1,2,2,2,4)^t$ for all $i=2, 3,4$.

\end{itemize}

At the end of the first cycle, the cost incurred due to maintenance of ${\bf M}_i$ when the system is critical is given by equation \eqref{eq:Costi}, i.e.
\[
Cost (1, \tau, i)=\widetilde{\alpha}_i(0) exp({\bf Q}_{i}^{*s}\tau){\bf D}_{1,U_2}({\bf M}_{i}\odot {\bf C}_{i})e, \quad i=1,2,3,4;
\]
and this expression is used to calculate the total cost incurred in the system maintenance action at time $\tau$ as it is expressed in Section  \ref{sec:maintpolicy}. Finally, the initial distribution of the $i$-th module for the next period is given by equation \ref {eq:alfai}
\[
\widetilde{\alpha_{i}}(1)=\widetilde{\alpha_{i}}(0)exp({\bf Q}_{i}^{*s}\tau){\bf M}_{i}, \quad i=1,2,3,4.
\]

Let us assume that the useful lifetime of the system (with maintenance) is denoted by $\T$, that is $\T=A \tau$, where $A$ is the maximum number of maintenance inspections carried out. The goal is to determine the optimal length of the operative cycles, i.e. $\tau$, in the sense of minimizing the total cost incurred by all maintenance actions implemented at the end of each $\tau$-cycle during the whole period $(0,\T]$. 

To select the optimal value of $\tau$, and consequently of $A$, given that duration of useful lifetime of the system is prespecified, we consider the following Montecarlo simulation  algorithm.

	\textbf{Algorithm}\
	
	\begin{enumerate}
		\item[1.] Define an equispaced grid of $M$ values $\{\tau_1, \ldots, \tau_M\}$ with $\tau_M=\mu)$; and set $m=0$;
		\item[2.] Put $m=m+1$ and:
		\begin{enumerate}
			\item[2.1.] Define $Cost^{(m)}$, an empty vectors of size $R$. Set $r=0$;
			\item[2.2.] Put $r=r+1$ and randomly simulate one path of the process in the interval $(0,\tau_m)$, that is $\{(\widetilde{i}_1,\widetilde{t}_1),\ldots,(\widetilde{i}_{n(\tau_m)},\widetilde{t}_{n(\tau_m)})\}$, where $\widetilde{i}_k$ are the successive visited states and $\widetilde{t}_{k}$ are the corresponding simulated jump times, so that $\widetilde{t}_{n(\tau_m)} \leq \tau_m$;
			\item[2.3.] Given $\widetilde{i}_{n(\tau_m)}$, consider the following cases:\
			\item If $\widetilde{i}_{n(\tau_m)} \in U_1$ then $Cost^{(m)}_r=C_I$;
			\item If $\widetilde{i}_{n(\tau_m)} \in U_2$ then $Cost^{(m)}_r=C_I+C_{M_1}+C_{M_2}+C_{M_3}+C_{M_4}$;
			\item If $\widetilde{i}_{n(\tau_m)} \in D$  then $Cost^{(m)}_r=C_I+\left(\tau_m-\widetilde{t}_{n(\tau_m)}\right)C_{Down}+C_{SR}$.
			
			\item[2.4.] Go to step 2.2 until $r=R$
		\end{enumerate}
		\item[3.] Define $Av\_cost^{(m)}=(1/R)\sum_{r=1}^R Cost^{(m)}_r$;
		\item[4.] Define $A^{(m)}=\mu/\tau_m$;
		\item[5.] Go to step 2 until $m=M$;
		\item[6.] Select $m_0= \underset{m}{\arg \min} \{Av\_cost^{(m)}A^{(m)}\}$, and fix $\tau=\tau_{m_0}$.
	\end{enumerate}

\begin{figure}[ht]
	\centering
	\includegraphics[width=0.6\textwidth]{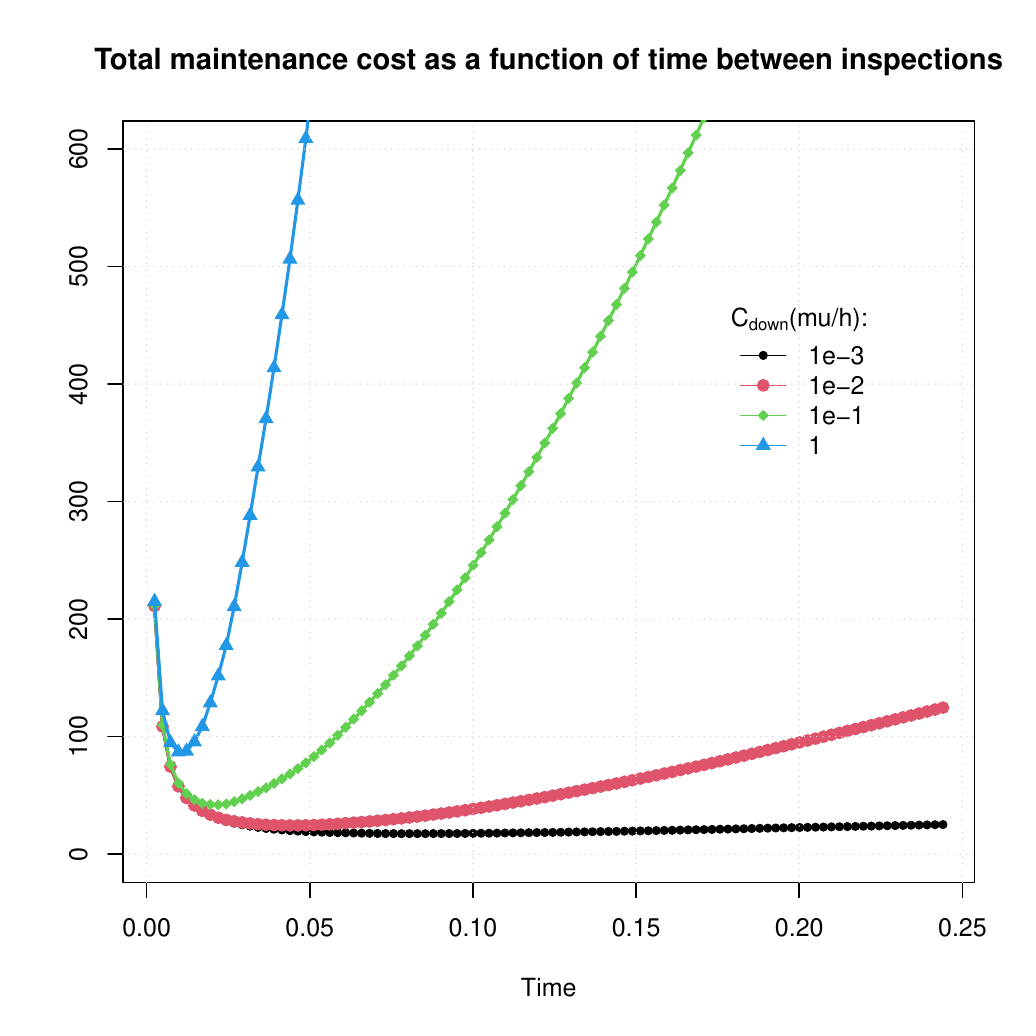}
	\vspace{-0.5cm}
	\caption{Maintenance strategy with 4 different specifications for $C_{down}$.}	\label{fig:4sc}
\end{figure}

\begin{figure}
	\centering
\includegraphics[width=0.45\textwidth]{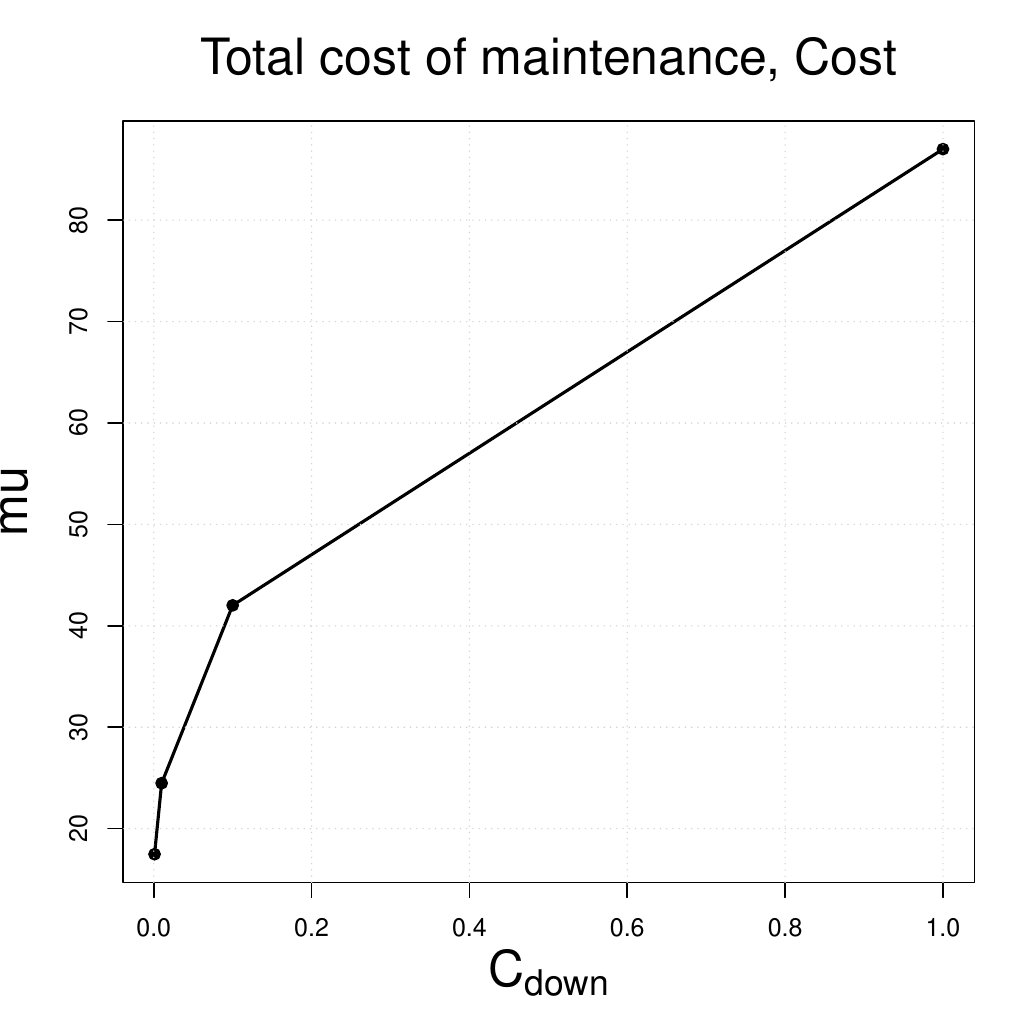}
\includegraphics[width=0.45\textwidth]{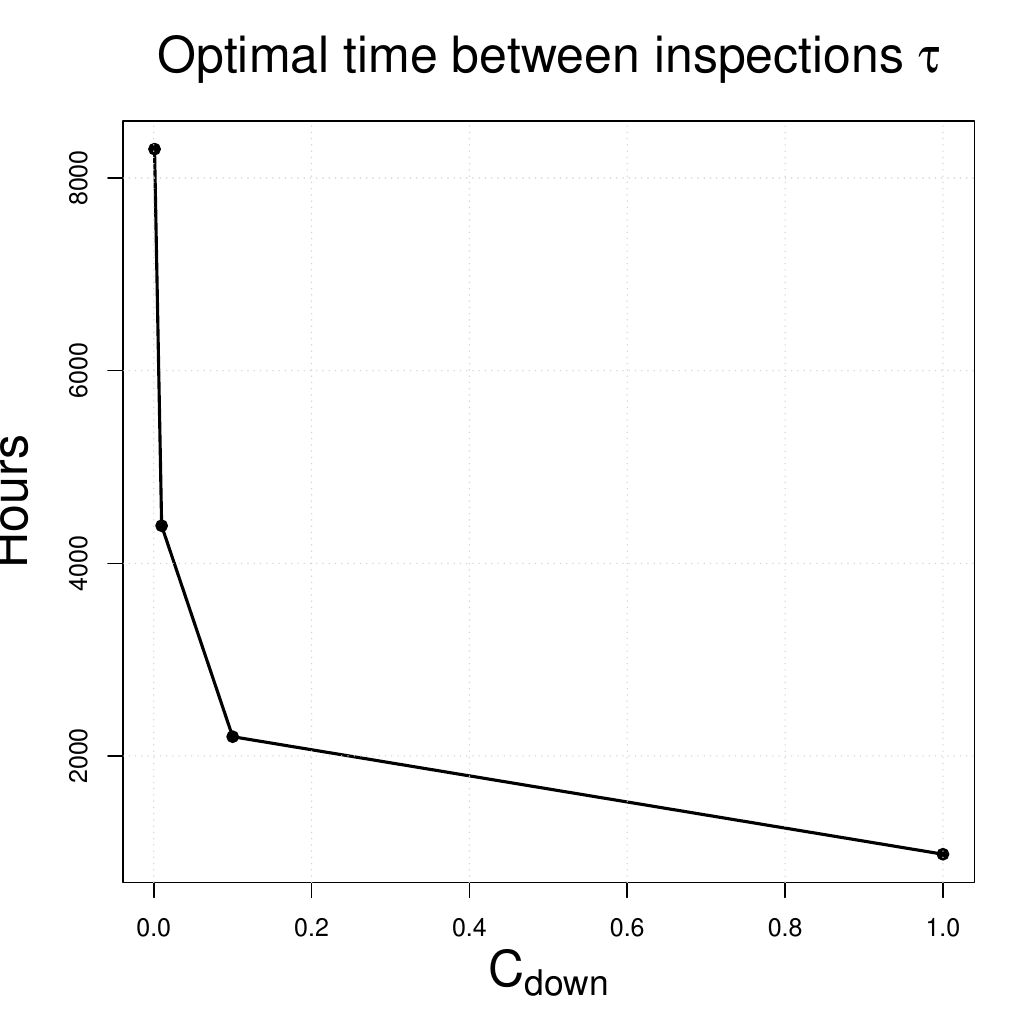}
\caption{{\it Left panel}: Total incurred cost of maintenance ($Cost$) for each value of non-operation cost.  {\it Right panel}: Optimal time of inspection ($\tau$) for each value of non-operation cost. }	\label{fig:summary}
\end{figure}

Figure \ref{fig:4sc} displays the results with the specifications of Table \ref{tab:cost}. The optimal values are given in Table \ref{tab:optimal}. Left panel of Figure \ref{fig:summary} displays the trend of total cost increase as the cost per unit time for non productivity of the system increases. Right panel of Figure \ref{fig:summary} displays the curve of optimal inspections times that is obtained for each scenario.

The computational time of this experiment is not an issue of significant concern even when a large number of iterations are considered; that is for a large value for $R$. For instance, a simulation with the specifications of Table \ref{tab:cost} for the first scenario (i.e. $c_{down}=0.001$, taking $R=50000$ and $M=100$) using a {\tt 11th Gen Intel(R) Core(TM) i7-1165G7 @ 2.80GHz 2.80 GHz} takes approximately 11.66 min.

\begin{table}[ht]
	\centering
	\begin{tabular}{|l||c|c|c|c|}
		\hline
		Scenario	& 1 & 2 & 3 & 4 \\ 		\hline \hline
		$C_{down}$ & 0.0010 & 0.0100 & 0.1000 & 1.0000 \\ \hline
		$\tau$ (\small \tt hours) & 8300 & 4390 &2200&980 \\ \hline
		$Cost \ ({\tt mu})$ & 17.4791 & 24.4786 & 42.0189 & 87.0051 \\ \hline
		$A$ & 6 & 11 & 23 & 51 \\		\hline
	\end{tabular}
	
	\caption{Optimal values considering 4 different scenarii. Note: ${\tt mu}$=monetary unit. }
	\label{tab:optimal}
\end{table}

\section{Conclusions}\label{sec:conclusions}
In the present paper we propose a maintenance policy for complex modular systems, where the modules in the system consist of a number of units. We assume that the system might fail due to the wear-out of its units or due to external shocks that represent changes in its environment. Additionally, the units lifetimes follow PH-distributions, where the phases represent different levels of degradation in the unit. The modeling of this system is based on MAM and it was addressed in a previous paper. 

Here we focus on the modeling of a maintenance policy for such a system, as well as the optimization of the cost associated to this policy, through the lifetime of the system. The proposed maintenance policy is a hierarchical decision-based policy where the system is inspected at regular intervals of time. At the time of inspection the system could be in an optimal, critical or failed state. If it is in an optimal state no maintenance action takes place. If it is in a failed state, the system is replaced. Finally, if it is in a critical state its modules are inspected. We check the individual modules for an optimal, critical or failed state. In this case, the same policy as for the system is applied, i.e. the module will be replaced if it is in a failed state. Alternatively, if it is in an optimal state no intervention will take place. If it is in a critical state, the state of units will be checked and the failed units will be replaced. 

For the proposed maintenance policy, we have defined maintenance matrices associated to the modules that describe the transition probabilities to an optimal state at inspection times, considering possible maintenance actions. Related to these matrices we have defined other matrices that provide the cost associated to the maintenance actions (described in the maintenance matrices). Additionally, we depict the stochastic process that describes the evolution of the system in time, considering performed maintenance actions. We have also provided the expressions to determine the initial probability vector, after every inspection, as well as the expected maintenance cost. From this, we can optimize the cost in terms of the frequency of inspections. 

The main contribution of this paper is the use of MAM to design a maintenance policy for complex modular systems. We have obtained closed-form expressions related to the proposed maintenance policy. The methodology here is applicable to other complex systems, based on layered composition/decomposition method. Finally, we have demonstrated the use of the methodology on a real use-case. Specifically we have considered the ECU system presented in \cite{Liu2014}, where no maintenance analysis  is carried out. This aspect is new in our study of this system. As we do not have true knowledge about the parameters involved in the cost analysis (i.e. $C_I$, $C_{RM}$, etc),  we have chosen all the values of Table \ref{tab:cost} which represent the different types of cost involved in our maintenance policy, arbitrarily and actually with no fully real meaning. Had we had expert guidance in this respect we could have built a meaningful maintenance-cost model, more interpretable from a practical decision-making sense in real life. Nevertheless, after running some simulations which are not shown in the paper, we have noticed that the most important index in our maintenance plan is the cost of non productivity per unit time, that is $C_{down}$. In other words, the parameter with respect to which the system is most sensitive is $C_{down}$, and then we have considered different scenarii according to the particular value of this parameter.\\

As future work, we want to incorporate dependency between the modules that are part of the system, presently our model considers that the modules fail independently. Additionally, we currently do not distinguish failure modes for the modules nor for the system. Therefore we would also consider different failure modes and related maintenance actions. Finally, we aim to use empirical data to demonstrate the proposed methodology.  
\section*{Acknowledgements}
 This work is jointly supported by the Spanish Ministry of Science and Innovation—State Research Agency through grants PID2020-120217RB-I00 and PID2021-123737NB-I00.



\begin{thebibliography}{}
	
	\bibitem[Drossel et~al.(2019)]{Drossel2019} Drossel WG , Zorn W, Hamm L. Modular system to measure and control force distribution in deep drawing processes to ensure part quality and process reliability. CIRP Ann 2019; 68 (1): 309--12. 
	  
	\bibitem[Shaik  et~al.(2015)]{Shaik2015} Shaik AM, Rao VK, Rao CS. Development of modular manufacturing systems – a review. Int J Adv Manuf Technol 2015;76 (5-8):789--802
	
	\bibitem[Debnath  et~al.(2014)]{Debnath2014} Debnath S, Qin J, Bahrani B, Saeedifard M, Barbosa P. Operation, control, and applications of the modular multilevel converter: A review. IEEE Trans Power Electron 2014;30(1):37--53.
	
	\bibitem[Alvarez  et~al.(2018)]{Alvarez2018} Alvarez EA, Rico-Secades M, Corominas EL, Herta-Medina N, Guitart JS. Design and control strategies for a modular hydrokinetic Smart grid. Int J Electr Power Energy Syst 2018;95:137--45.
	
	\bibitem[Poudel  et~al.(2019)]{Poudel2019} Poudel B, Joshi K, Gokaraju R. A dynamic model of small modular reactor based nuclear plant for power system studies. IEEE Trans Energy Convers 2019;35(2) 977--85. 
	
	\bibitem[Altaee  et~al.(2019)]{Altaee2019} Altaee A, Cipolina A. Modelling and optimization of modular system for power generation from salinity gradient. Renewable Energy 2019; 201( 141):139—47.
	
	\bibitem[Cai  et~al.(2023)]{Cai2023} G. Cai, Y. Huang, B. Chen, Y. Shen, X. Shi, B. Peng, S. Mi, J. Huang. Modular design of centrifugal microfluidic system and its application in nucleic acid screening. Taranta 259 (2023) 124486. 
	
	\bibitem[Shaik  et~al.(2015)]{Abdul2015} Shaik AM,  Rao VVSK, Rao CS. Development of modular manufacturing systems- a review. Int J Adv Manuf Technol 2015; 76:789—802.
	
	\bibitem[Xiahou et~al.(2023)]{TangFan2023} Xiahou T,  Zheng YX, Liu Y, Chen H. Reliability modeling of modular k-out-ot-n systems with functional dependency: A case of study of radar transmitter systems. Reliab Eng Syst Saf 2023; 233, 109120.
	
	\bibitem[Putri and Buana (2020)]{Nilda2020} Putri NT, Buana FS. Preventive Maintenance Scheduling by Modularity Design Applied to Limestone Crusher Machine. Procedia Manufacturing  2020; 43 682--687.
	
	\bibitem[Bonvoisin et~al.(2016) ]{Jeremy2016} Bonvoisin J, Halstenberg F, Buchert T, Stark RG. A systematic literature review on modular product design. Journal of Engineering Design  2016; 27 (7) 1--27.
	
	\bibitem[Baldwin and Clark(2020)]{Baldwin2000} Baldwin CY and Clark KB. Design rules: The Power of Modularity Cambridge MA: MIT Press 2020. 
	
	\bibitem[Li et~al.(2021)]{Li2021} Li, H.; Díaz, H.; Guedes Soares, C. A developed failure mode and effect analysis for floating offshore wind turbine support structures. { Renew. Energy} { 2021}; {164}, 133--145. 
	
	\bibitem[Dorronsoro(2022)]{Dorronsoro2022} Dorronsoro X, Garayalde E, Iraola U, Aizpurua M. Modular battery energy storage system design factors analysis to improve battery-pack reliability. Journal of Energy storage 2022; 54, 105256.  
	
	\bibitem[Yicong(2016)]{Yicong2016} Yicong G, Yixiong F, Jianrong T. Product modular design incorporating preventive maintenance issues. Chinese Journal of Mechanical Engineering 2016; 29, 406--420. 
		
	\bibitem[Sharifi and Taghipour(2020)]{Sharifi2020} Mani Sharifi, Taghipour S. Optimal inspection interval fpr a k-out-of-n system with non-identical components. Journal of Manufacturing System 2020; 55, 233--247. 
	

	\bibitem[Gamiz et~al.(2022)]{Gamiz2022} Gamiz, ML;  Limnios, N,  Segovia-Garcia, MC;  Hidden Markov Models in Reliability. Eur J Oper Res 2023; 304(3), 1242--1255.

	
	\bibitem[Wang et~al.(2019)]{Wang2019} Wang J., Makis V., Zhao X. Optimal condition-based and age-based opportunistic maintenance policy for a two-unit series system. Comput Ind Eng, 2019; 134,  1--10.
	
	\bibitem[Cavalcante et~al.(2021)]{Cavalcante2021} Cavalcante C, Lopes R, Scarf P. Inspection and replacement policy with a fixed periodic Schedule. Reliab Eng Syst Saf 2021; 208, 107402.
		
	\bibitem[Zhang et~al.(2023)]{Zhang2023} Zhang N, Deng Y, Liu B, Zhang J. Condition-based maintenance for a multi-component system in a dynamic operating environment.  Reliab Eng Syst Saf 2023; 231, 108988.
	
	\bibitem[Caballe et~al.(2015)]{Caballe2015} Caballé NC, Castro IT, Pérez CJ,  Lanza-Gutiérrez JM. A condition-based maintenance of a dependent degradation-threshold-shock model in a system with multiple degradation process. Reliab Eng Syst Saf 2015; 134, 98--109.
	
	\bibitem[Hashemi et~al.(2020)]{Hashemi2020} Hashemi M, Asadi M, Zarezadeh S. Optimal maintenance policies for coherent systems with multi-type components.  Reliab Eng Syst Saf 2020; 195, 106674.
	
	\bibitem[Zhao et~al.(2020)]{Zhao2020} Zhao J, Si S, Cai Z,  Guo P,  Zhu W. Mission sucess probabilitly optimization for phased-mission systems with repairable component modules.  Reliab Eng Syst Saf 2020;195, 106750.  
	
	\bibitem[Li(2016)]{Li2016}Li J. Reliability calculation of a parallel redundant system with different failure rate \& repair rate using Markov modeling. Journal of Reliability and Statistical Studies 2016; 9(1), 1-10. 
	
	\bibitem[Liu et~al.(2014)]{Liu2014} Liu Z, Liu Y, Cai B. Reliability Analysis of the Electrical Control System of Subsea Blowout Preventers Using Markov Models. { PLoS ONE} { 2014}; 9, 1--9. 	
	
	\bibitem[Gamiz et~al.(2022)]{MOMA2022} Gámiz ML, Montoro-Cazorla D, Segovia-García MC, Pérez-Ocón, R. {\it MoMA} Algorithm: A Bottom-Up Modeling Procedure for a Modular System under Environmental Conditions. Mathematics 2022, 10, 3521. 
	
	\bibitem[Chen et~al.(2022)]{Chen2022} Chen Y, Qiu Q, Zhao X. Condition-based opportunistic maintenance policies with two-phase inspections for continuous-state systems.  Reliab Eng Syst Saf 2022; 228, 108767.
	
	\bibitem[Montoro-Cazorla and Perez-Ocon(2015)]{Montoro2015} Montoro-Cazorla D, P\'{e}rez-Oc\'{o}n R. A reliability system under cumulative shocks governed by a BMAP. { Appl. Math. Model.} {2015}; { 39},  7620--7629.
		
	\bibitem[Segovia and Labeau(2013)]{Segovia2013} Segovia MC,  Labeau PE. Reliability of a multi-state system subject to shocks using phase-type distributions. { Appl. Math. Model.} { 2013}; { 37}, 4883--4904.
	
	\bibitem[Montoro-Cazorla and Perez-Ocon(2018)]{Montoro2018} Montoro-Cazorla D, P\'{e}rez-Oc\'{o}n R. Constructing a Markov process for modelling a reliability system under multiple failures and replacements.   Reliab Eng Syst Saf  { 2018}; { 173}, 34--47.
	
	\bibitem[Chakravarthy(2012)]{Chakravarthy2012} Chakravarthy SR. Maintenance of a deteriorating single server system with Markovian arrivals and random shocks.  { Eur J Oper Res} { 2012}; { 222}, 508--522.
	
	
	\bibitem[Maaroufi et~al.(2013)]{Maaroufi2013}  Maaroufi G, Chelbi A,  Rezg N. Optimal selective renewal policy for systems subject to propagated failures with global effect and failure isolation phenomena.  Reliab Eng Syst Saf 2013; 114, 61--70. 
	
	\bibitem[Levi et~al.(2014)]{Levi2014} Levi R, Magnanti T, Muckstadt J, Segev D. and Zarybnisky E. Maintenance Sheduling for Modular Systems: Modeling and Algorithms. Naval Research Logistics 2014; 61.
	
	\bibitem[Joo(2009)]{SeongJong2009}  Joo SJ. Sheduling preventive maintenance for modular designed components: A dynamic approach. Eur J Oper Res 2009; 192, 512--520. 
	
	\bibitem[Karmarkar and Kubat(1987)]{Uday1987} Karmarkar US, Kubat P. Modular product design and product support. Eur J Oper Res 1987; 291, 74--82. 
	
	\bibitem[Bellman(1970)]{Bellman} Bellman R.  { Introduction to Matrix Analysis}. McGraw-Hill: New York, NY, USA,	{ 1970}.
	
	\bibitem[Graham(1981)]{Graham} Graham A.  {	Kronecker Products, and Matrix Calculus with Applications}.  Halsted Press: Chichester, UK;  Horwood, NY, USA, {1981}.
	
	\bibitem[Neuts(1981)]{Neuts1981} Neuts MF. { Matrix-Geometric Solutions in Stochastic Models: An Algorithm Approach}; John Hopkins University Press: Baltimore, Maryland, {1981}.
	

	\bibitem[Torrado et~al.(2021)]{Torrado21} Torrado N, Arriaza A, Navarro J. A study on multi-level redundancy allocation in coherent systems formed by modules Reliab Eng Syst Saf 2021; 213, 107694. 
	
	
	\bibitem[Liang et~al.(2022)]{Qingzhu2022} Liang Q, Yang Y, Zhang H, Peng C,  Lu J. . Analysis of simplification in Markov state-based models for reliability assessment of complex safety systems. Reliab Eng Syst Saf 2022; 221, 108373.
	
	
\end{thebibliography}
\end{document}